\documentclass[12pt]{article}

\ifx\pdfoutput\undefined
\usepackage[dvips,bookmarks]{hyperref}
\else
\usepackage{hyperref}
\fi
\hypersetup{colorlinks=false,bookmarksopen,bookmarksnumbered,citecolor=blue,
   pdfstartview=FitH}

\usepackage{latexsym}
\usepackage{amssymb,amsfonts,amsmath}
\usepackage{graphicx} 
\usepackage{indentfirst}
\usepackage{bbm}
\usepackage{amssymb}
\usepackage{verbatim}
\usepackage{amsmath, amsthm,amssymb}
\usepackage{mathrsfs}
\usepackage{hyperref}
\usepackage{amsfonts}
\usepackage{dsfont}
\usepackage{slashed, tensor}
\usepackage{booktabs}
\usepackage{graphicx}
\usepackage{mathrsfs}

\usepackage[colorinlistoftodos,prependcaption,textsize=small]{todonotes}
\parskip=\medskipamount

\newcommand {\cD}{{\cal D}}

\newcommand {\cG}{{\cal G}}
\newcommand {\cH}{{\cal H}}
\newcommand {\cI}{{\cal I}}

\newcommand {\cK}{{\cal K}}
\newcommand {\cL}{{\cal L}}
\newcommand {\cM}{{\cal M}}
\newcommand {\cN}{{\cal N}}
\newcommand {\cO}{{\cal O}}

\newcommand {\cR}{{\cal R}}
\newcommand {\cS}{{\cal S}}
\newcommand {\cT}{{\cal T}}

\newcommand {\cV}{{\cal V}}
\newcommand {\cW}{{\cal W}}

\def\a{\alpha}
\def\b{\beta}
\def\c{\chi}
\def\d{\delta}

\def\g{\gamma}

\def\k{\kappa}
\def\l{\lambda}
\def\m{\mu}

\def\o{\omega}

\def\q{\theta}
\def\r{\rho}
\def\s{\sigma}
\def\t{\tau}

\def\D{\Delta}
\def\F{\Phi}

\def\L{\Lambda}
\def\O{\Omega}

\def\U{\Upsilon}

\def\ri{{\rm i}}
\def\re{{\rm e}}

\newcommand{\ve}{\varepsilon}                            

\newcommand{\pa}{\partial}                           
\newcommand{\hf}{\frac12}

\newcommand{\vf}{\varphi}

\newcommand{\be}{\begin{equation}}
\newcommand{\ee}{\end{equation}}
\newcommand{\bea}{\begin{eqnarray}}
\newcommand{\eea}{\end{eqnarray}}
\newcommand{\non}{\nonumber}
\newcommand{\ba}{\begin{array}}
\newcommand{\ea}{\end{array}}

\newcommand{\dsC}{{\mathbb C}}

\newcommand{\bm}[1]{\mbox{\boldmath$#1$}}

\def\double #1{#1{\hbox{\kern-2pt $#1$}}}

\newcommand{\bsubeq}{\begin{subequations}}
\newcommand{\esubeq}{\end{subequations}}

\newcommand{\ul}{\underline}
\newcommand{\eps}{{\ve}}

\newcommand{\rd}{\mathrm d}
\numberwithin{equation}{section}

\newcommand{\RM}{R(M)}
\newcommand{\RD}{R(\mathbb D)}
\newcommand{\RN}{R(N)}

\newcommand{\RS}{R(S)}
\newcommand{\RK}{R(K)}

\newcommand{\bai}{{\bar i}}
\newcommand{\baj}{{\bar j}}
\newcommand{\bak}{{\bar k}}
\newcommand{\bal}{{\bar l}}

\newcommand{\bau}{{\bar 1}}

\newcommand{\rL}{{\rm L}}
\newcommand{\rR}{{\rm R}}

\begin{document}

\begin{titlepage}
\begin{flushright}
October, 2016 \\
\end{flushright}
\vspace{5mm}

\begin{center}
{\Large \bf 
Minimal \mbox{$\bm{\cN = 4}$} topologically massive supergravity
}
\\ 
\end{center}

\begin{center}

{\bf
Sergei M. Kuzenko${}^{a}$, Joseph Novak${}^{b}$ and Ivo Sachs${}^{c}$
} \\
\vspace{5mm}

\footnotesize{
${}^{a}${\it School of Physics M013, The University of Western Australia\\
35 Stirling Highway, Crawley W.A. 6009, Australia}}  
~\\
\vspace{2mm}
\footnotesize{
${}^{b}${\it Max-Planck-Institut f\"ur Gravitationsphysik, Albert-Einstein-Institut\\
Am M\"uhlenberg 1, D-14476 Golm, Germany}
}
~\\
\vspace{2mm}
\footnotesize{
${}^{c}${\it Arnold Sommerfeld Center for Theoretical Physics, Ludwig-Maximilians-Universit\"at\\
Theresienstra§e 37, D-80333 M\"unchen, Germany}
}
~\\
\vspace{2mm}
\texttt{sergei.kuzenko@uwa.edu.au, joseph.novak@aei.mpg.de, ivo.sachs@physik.uni-muenchen.de}\\
\vspace{2mm}

\end{center}

\begin{abstract}
\baselineskip=14pt
Using the superconformal framework, we construct a new off-shell model for $\cN=4$ topologically massive supergravity which is minimal in the sense that it makes use of a single compensating vector multiplet and involves no free parameter.
As such, it provides a counterexample to the common lore that two compensating multiplets are required
within the conformal approach to supergravity with eight supercharges in diverse dimensions. This  theory is an off-shell $\cN=4$ supersymmetric extension of 
chiral gravity. 
All of its solutions correspond to non-conformally flat superspaces.
Its maximally supersymmetric solutions 
include the so-called critical (4,0) anti-de Sitter superspace introduced in arXiv:1205.4622, and well as 
warped critical (4,0) anti-de Sitter superspaces. 
We also propose a dual formulation for the theory in which the vector multiplet 
is replaced with an off-shell hypermultiplet. Upon elimination of the auxiliary 
fields belonging to the hypermultiplet and imposing certain gauge conditions, 
the dual action reduces to the one introduced in arXiv:1605.00103.
\end{abstract}

\vfill

\vfill
\end{titlepage}

\newpage
\renewcommand{\thefootnote}{\arabic{footnote}}
\setcounter{footnote}{0}


\allowdisplaybreaks


\section{Introduction}

A unique feature of three spacetime dimensions (3D) is the existence of topologically 
massive Yang-Mills and gravity theories. They are obtained by augmenting 
the usual Yang-Mills action or the gravitational action 
by a  gauge-invariant topological mass term.
Such a mass term coincides with a non-Abelian Chern-Simons action in the Yang-Mills case \cite{Siegel,Schonfeld,DJT1,DJT2} 
and with a Lorentzian Chern-Simons term in the case of gravity \cite{DJT1,DJT2}. Without adding the Lorentzian Chern-Simons term, 
the pure gravity action propagates no local degrees of freedom. The Lorentzian Chern-Simons term can be interpreted 
as the action for conformal gravity
in three dimensions \cite{DJT1,vN,HW}.\footnote{The usual Einstein-Hilbert
action for 3D gravity with a cosmological term can also be interpreted as the Chern-Simons 
action for the anti-de Sitter group \cite{AT,Witten}.}

Topologically massive theories of gravity possess supersymmetric extensions.
In particular, $\cN=1$ topologically massive supergravity 
was introduced in \cite{DK} and its cosmological extension followed in \cite{Deser}.
The off-shell formulations for $\cN$-extended topologically massive supergravity theories were presented 
in \cite{KLRST-M} for $\cN=2$ and in  \cite{KN14} for $\cN=3$ and $\cN=4$.  
In all of these theories, the action functional is a sum of two terms, 
one of which is the action for pure $\cN$-extended supergravity 
(Poincar\'e or anti-de Sitter)
and the other is the action for $\cN$-extended conformal supergravity.
The off-shell actions for $\cN$-extended supergravity theories 
in three dimensions were given
in \cite{GGRS} for $\cN=1$, \cite{KLT-M11,KT-M11} for $\cN=2$, 
and \cite{KLT-M11} for the cases $\cN=3, ~4$.
The off-shell actions for $\cN$-extended conformal supergravity 
were given in \cite{vN} for $\cN=1$, \cite{RvanN86} for $\cN=2$, and 
 \cite{BKNT-M2} for 
 $\cN=3, ~4$. The latter work made use of
 the formulation for $\cN$-extended conformal supergravity 
 presented in \cite{BKNT-M1}.
 
 The off-shell structure of 3D $\cN=4$ supergravity \cite{KLT-M11} is analogous 
 to that of 4D $\cN=2$ supergravity (see, e.g., \cite{FVP} for a pedagogical review) 
 in the sense that  two superconformal compensators
 are required (for instance, two off-shell vector multiplets, 
 one of which is self-dual and the other anti-self-dual)
 in order to realise pure Poincar\'e or anti-de Sitter (AdS) supergravity theories.
 We recall that the equations of motion for pure $\cN=4$ Poincar\'e or AdS supergravity
 are inconsistent if one makes use of a single compensator \cite{KN14}. 
 By construction, the off-shell $\cN=4$ topologically massive supergravity theory of \cite{KN14} makes use of two compensators.
 However, in \cite{LS16} the consistent system of dynamical equations was proposed
for $\cN=4$ topologically massive AdS supergravity with a single compensating  hypermultiplet, 
 following earlier work in \cite{Chu:2009gi,Gran:2012mg,Nilsson:2013fya} on ABJ(M) models. 
 A peculiar feature of this model, like those considered in \cite{Chu:2009gi,Gran:2012mg,Nilsson:2013fya}, is that it has no free parameter. Consequently the dimensionless combination, $\mu\ell$, of mass $\m$ and AdS radius $\ell$
 takes a fixed value, $\mu\ell=1$, as in chiral gravity \cite{LSS}.
 In \cite{LSS} it was argued that $\mu\ell=1$ is the only value for the quantum theory 
 to have a chance to be free of ghosts. 
It is thus interesting that the $\cN=4$ theory of \cite{LS16} picks precisely this value.\footnote{The only 
 known models which pick precisely this value are the topologically gauged ABJ(M) models
of  \cite{Chu:2009gi,Gran:2012mg,Nilsson:2013fya}.}  
 
 In \cite{LS16} a supergravity action functional 
was also postulated to generate the dynamical equations given.
This action was claimed to be off-shell without giving technical details. 
 In this paper we propose a new off-shell model 
 for $\cN=4$ topologically massive supergravity 
which is minimal in the sense that it makes use of a single compensating 
vector multiplet. 
 The theory is consistent only if the term corresponding to $\cN=4$ conformal supergravity 
 is turned on. An important maximally supersymmetric solution for this theory is
the so-called critical (4,0) AdS superspace introduced in \cite{KLT-M12}.
Our supergravity theory is first presented  in a manifestly supersymmetric form, 
and then its action functional is reduced to components. 
By choosing  appropriate gauge conditions at the component level and performing a duality transformation, we show how to reduce our off-shell supergravity action to the one postulated in \cite{LS16}.

This paper is organised as follows. In section 2 we recall the superspace 
geometry of the two $\cN=4$ vector multiplets 
and the corresponding locally supersymmetric actions.
In section 3 we present two models for minimal $\cN=4$ topologically massive 
supergravity, analyse their equations of motion and give
a brief discussion of the maximally supersymmetric solutions.
 Section 4 is devoted to the component structure of minimal $\cN=4$ topologically massive 
 supergravity. Concluding comments are given in section 5.
The main body of the paper is accompanied with three technical appendices.
The essential details of the known superspace formulations for $\cN=4$ conformal
supergravity are collected in Appendices A and B. 
Some useful super-Weyl gauge conditions in 
SO(4) superspace and their implications are given in Appendix C.


\section{The $\cN = 4$ vector multiplets}\label{VM}

There are two inequivalent irreducible $\cN=4$ vector multiplets in three dimensions, 
self-dual and anti-self-dual ones, as  discovered by Brooks and Gates  \cite{BrooksG}.
In this section we review the superspace 
geometry of these supermultiplets in the presence of $\cN=4$ conformal supergravity 
\cite{KLT-M11,BKNT-M1}
and the corresponding locally supersymmetric actions \cite{KLT-M11}.

Throughout this paper we make use of both the SO(4) superspace formulation 
of conformal supergravity, which was sketched in \cite{HIPT}
and fully developed  in \cite{KLT-M11}, 
and the conformal superspace formulation presented in \cite{BKNT-M1}. 
These formulations are related to each other since SO(4) superspace may be viewed 
as a gauge fixed version of the $\cN = 4$ conformal 
superspace \cite{BKNT-M1}. Due to this reason,
we will first start by formulating vector multiplets in conformal superspace. 
We refer the reader to Appendix \ref{geometry} for the salient details of the 
conformal superspace formulation. The geometry of SO(4) superspace 
in briefly reviewed in Appendix \ref{geometrySO4}.

\subsection{Kinematics}

To describe an Abelian vector multiplet in a curved superspace $\cM^{3|8}$ 
 parametrised by coordinates $z^M = (x^m, \ \q^\mu_\cI)$, 
we introduce gauge covariant derivatives
\bea
\bm \nabla = E^A \bm \nabla_A \ , \quad 
{\bm\nabla}_A =({\bm\nabla}_a, {\bm\nabla}_\a^I)
:= \nabla_A - V_A \bm Z~, \quad [\bm Z , \bm \nabla_A] = 0 \ ,
\eea
with $E^A = \rd Z^M E_M{}^A $ the superspace vielbein, 
$\nabla_A$ the superspace covariant derivatives \eqref{A.2}
obeying the (anti-)commutation relations \eqref{nablanabla}, and 
$V = E^A V_A$ the gauge connection associated with the generator $\bm Z$. 
The gauge transformation of $V$ is
\be 
\d V = \rd \t  \ ,
\label{2.2}
\ee
where the gauge parameter $\t (z)$ is an arbitrary real scalar superfield.

The algebra of gauge covariant derivatives is
\begin{align} [{\bm \nabla}_A, {\bm \nabla}_B\} &= -T_{AB}{}^C{\bm \nabla}_C
-\hf \RM_{AB}{}^{cd} M_{cd}
-\hf \RN_{AB}{}^{PQ} N_{PQ}
- \RD_{AB} \mathbb D \non\\
&\quad - \RS_{AB}{}^\g_I S_\g^I
	- \RK_{AB}{}^c K_c
	- F_{AB} \bm Z \ ,
\end{align}
where the torsion and curvatures are those of conformal superspace but with $F_{AB}$ corresponding 
to the gauge covariant field strength $F = \hf E^B \wedge E^A F_{AB} = \rd V$.
The field strength $F_{AB}$ 
satisfies the Bianchi identity
\be \rd F = 0 \ , \quad \nabla_{[A} F_{BC\}} + T_{[AB}{}^D F_{|D| C\}} = 0
\ee
and must be subject to covariant constraints to describe an irreducible vector multiplet.

In order to describe an $\cN=4$ vector multiplet,
the superform $F$ is subject to the constraint
 (see \cite{KLT-M11} for more details)
\bsubeq
\bea
F_{\a}^I{}_\b^J &=&  -2\ri\ve_{\a\b}G^{IJ} \ , \quad G^{IJ}=-G^{JI}~,
\eea
and then the Bianchi identity fixes the remaining components of $F$
 to be 
\bea
F_{a}{}_\b^J&=&
\frac{1}{ 3}(\g_a)_\b{}^{\g} \nabla_{\g K} G^{JK}
~,
\\
F_{ab}&=&
-\frac{\ri}{ 48}\ve_{abc}(\g^c)^{\a\b}[\nabla_{\a}^{ K}, \nabla_{\b}^{ L}] G_{ K L}
~,
\eea
\esubeq
where $G^{IJ}$ is 
primary and of dimension $1$,
\be
S_\a^I G^{JK} = 0 \ , \quad K_a G^{IJ}=0\ , 
\quad \mathbb D G^{IJ} = G^{IJ} \ .
\ee
Moreover, the field strength $G^{I J}$ 
is constrained by the dimension-3/2 Bianchi identity 
\bea
\nabla_{\g}^{I} G^{ J K}&=&
\nabla_{\g}^{[I} G^{ J K]}
- \frac{2}{ 3} \d^{I [J} \nabla_{\g L} G^{ K] L}
~.
\label{VMBI}
\eea

It is well known (see \cite{KLT-M11} and references therein) 
that the constraint \eqref{VMBI} defines a reducible 
off-shell supermultiplet.\footnote{Such a long $\cN=4$ supermultiplet naturally
originates upon reduction of any off-shell $\cN > 4$ vector multiplet to $\cN = 4$
superspace \cite{KS}.} 
The point is that 
 the Hodge-dual of $G^{IJ}$, 
\be
\tilde{G}^{IJ} := \hf \ve^{IJKL}G_{KL} \ , 
\label{4.17}
\ee
obeys the same constraint as $G^{IJ}$ does, 
\begin{subequations} \label{4.18a}
\bea
\nabla_{\g}^{I} \tilde{G}^{ J K}&=&
\nabla_{\g}^{[I} \tilde{G}^{ J K]}
- \frac{2}{3} \d^{I [J} \nabla_{\g L} \tilde{G}^{ K] L} \ ,
\eea
\end{subequations} 
where $\eps^{IJKL}$ is the Levi-Civita
tensor.
As a result one may constrain 
the field strength $G^{IJ}$ to be self-dual, $\tilde{G}^{IJ} = {G}^{IJ}$
or anti-self-dual, $\tilde{G}^{IJ} = -{G}^{IJ}$.
These choices correspond to two different irreducible off-shell $\cN=4$ vector multiplets, which we denote by $G_+^{IJ}$ and $G_-^{IJ}$, 
respectively.
In what follows we will make use of an (anti-)self-dual Abelian vector multiplet 
such that its field strength $G^{IJ}_{\pm}$ is nowhere vanishing, 
$G_{\pm}^2 := \hf G_{\pm}^{IJ} G_{\pm IJ} \neq 0$.

When working with $\cN=4$ supersymmetric theories, 
a powerful technical tool is
the isospinor notation based on the isomorphism 
 ${\rm SO}(4) \cong  \big( {\rm SU}(2)_{\rL}\times {\rm SU}(2)_{\rR}\big)/{\mathbb Z}_2$, 
which allows one to replace each SO(4) vector 
index  with a pair of isospinor ones. 
In defining the isospinor notation, we follow \cite{KLT-M11} 
and associate with a real SO(4) vector $V_I$ a second-rank isospinor $V_{i\bar{i}}$ 
defined as
\be V_I \rightarrow V_{i\bar{i}} := (\t^I)_{i\bar{i}} V_I \ , \quad V_I = \t_I{}^{i\bar{i}} V_{i\bar{i}} \ , \quad (V_{i\bar{i}})^* = V^{i\bar{i}}\, ,
\ee
where we have introduced the $\t$-matrices
\bea
(\t^I)_{i\bar{i}}=({\mathbbm 1},\ri\s_1,\ri\s_2,\ri\s_3)
~,~~~~~~
I={ 1},\cdots,{ 4}~,~~~
i=1,2~,~~
\bar{i}=\bar{1},\bar{2}\ .
\eea
The isospinor indices of $\rm SU(2)_L$ and $\rm SU(2)_R$ spinors $\psi_i$ and $\c_{\bar{i}}$, respectively, are
raised and lowered using 
the antisymmetric  
tensors $\ve^{ij},\ve_{ij}$ and $\ve^{{\bar i}{\bar j}},\ve_{{\bar i}{\bar j}}$
(normalised by $\ve^{12}=\ve_{21}=\ve^{{\bar 1}{\bar 2}}=\ve_{{\bar 2}{\bar 1}}=1$) 
according to
\be \psi^i = \eps^{ij} \psi_j \ , \quad \psi_i = \eps_{ij} \psi^j \ , 
\quad \c^{\bar{i}} = \eps^{\bar{i}\bar{j}} \c_{\bar{j}} \ , 
\quad \c_{\bar{i}} = \eps_{\bar{i}\bar{j}} \c^{\bar{j}} \ .
\ee
We then have the following dictionary:
\bsubeq
\bea
V^I U_I &=& V^{i\bar{i}} U_{i\bar{i}} \ , \\
A_{i\bar{i}j\bar{j}} := A_{IJ} (\t^I)_{i\bar{i}}  (\t^J)_{j\bar{j}} 
&=& \eps_{ij} A_{\bar{i}\bar{j}} + \eps_{\bar{i}\bar{j}} A_{ij} \ , 
\quad A_{ij} = A_{ji} \ , \quad A_{\bar{i}\bar{j}} = A_{\bar{j}\bar{i}} \ , \\
\hf A^{IJ} B_{IJ} &=& A^{ij} B_{ij} + A^{\bar{i}\bar{j}} B_{\bar{i}\bar{j}} \ , \\
\eps_{i\bar{i}j\bar{j}k\bar{k}l\bar{l}} &=& \eps_{ij} \eps_{kl} \eps_{\bar{i}\bar{l}} \eps_{\bar{j}\bar{k}}
- \eps_{il} \eps_{jk} \eps_{\bar{i}\bar{j}} \eps_{\bar{k}\bar{l}} \ ,
\label{213d}
\eea
\esubeq
where $V^I$ and $U^I$ are SO(4) vectors, $A^{IJ}$ and $B^{IJ}$ are anti-symmetric second-rank SO(4) tensors. The left-hand side of  \eqref{213d} is
the Levi-Civita tensor in the isospinor notation.

In the isospinor notation, the self-dual ($G^{IJ}_+$) and  anti-self-dual 
($G^{IJ}_-$) vector multiplets take the form
\bea
G_+^{\,i\bar{i}j\bar{j}} = -  \eps^{ij} G^{\bar{i}\bar{j}} \,, \qquad 
G_-^{\,i\bar{i}j\bar{j}} =- \eps^{\bar{i}\bar{j}} G^{ij} \, ,
\eea
and the Bianchi identity \eqref{VMBI} turns into
\bea
\nabla_\a^{(i\bai}G^{kl)}=0\, ,\qquad
\nabla_\a^{i(\bai}G^{\bak\bal)}=0 \,.
\label{VMBI2} 
\eea
At this stage it is useful to introduce left and right isospinor variables $v_\rL := v^i \in \dsC^2 \setminus \{0\}$ and $v_\rR := v^{\bar{i}} \in \dsC^2 \setminus \{0\}$,
which can be used to package the anti-self-dual field strength $G^{ij}$ 
and the self-dual field strength $G^{\bar{i}\bar{j}}$
into fields without isospinor indices, $G_{\rm L}^{(2)} (v_\rL) := G_{ij} v^i v^j$ and 
$G_{\rm R}^{(2)} (v_\rR) := G_{\bar i \bar j} v^{\bar i} v^{\bar j}$, 
respectively.
The same isospinor variables can be used
to define two different subsets,  
$\nabla_\a^{(1)\bai}$ and $\nabla_\a^{(\bau)i}$,
in the set of spinor covariant  derivatives $\nabla^{i\bar i}_\a$ by the rule
\bea
\nabla_\a^{(1)\bai}:=v_i\nabla_\a^{i\bai}~,~~~~~~
\nabla_\a^{(\bau)i}:=v_\bai\nabla_\a^{i\bai}~.
\eea
It follows from \eqref{A.33} that the operators $\nabla_\a^{(1)\bar i}$
obey the anti-commutation relations:
\bea
\big\{ \nabla_\a^{(1)\bar i} , \nabla_\b^{(1)\bar j} \big\} &=& 2 \ri \eps_{\a\b} \eps^{\bar{i} \bar{j}} W 
{ L}^{(2)} 
+ \ri \eps_{\a\b} \eps^{\bar{i} \bar{j}} \nabla^{\g (1)}{}_{\bar{k}} W S_\g^{(1)\bar{k}} \non\\
&&- \frac{1}{4} \eps_{\a\b} \eps^{\bar{i} \bar{j}} \nabla_\g{}^{(1)}{}_{\bar{k}} \nabla_\d^{(1) \bar{k}} W K^{\g\d} \ ,
\label{217}
\eea
where $L^{(2)} = v_i v_j L^{ij}$ and $S_\a^{(1)\bar{i}} $ is defined similarly to   $\nabla_\a^{(1)\bar i}$.
The rationale for the definitions given is that 
the constraints \eqref{VMBI2} now become the analyticity conditions
\bea
\nabla_\a^{(1)\bai}G_\rL^{(2)}&=&0~,  \qquad
\nabla_\a^{(\bau)i}G_\rR^{(2)}=0~.
\label{VMBI3} 
\eea
which tell us that each of $G_\rL^{(2)}$ and $G_\rR^{(2)}$ depends 
on half  the Grassmann coordinates. The constraints \eqref{VMBI3}  do not change 
under re-scalings $v^i \to c_\rL v^i$ and $v^{\bar i} \to c_\rR v^{\bar i}$, 
with $c_\rL, \, c_\rR \in \dsC \setminus \{0\}$, with respect to which 
$G_\rL^{(2)}(v_\rL)$ and $G_\rR^{(2)}(v_\rR)$ are homogeneous polynomials
of degree two. We see that the isospinor variables $v_\rL$ and $v_\rR$ are defined
modulo the equivalence relations 
$v^i \sim c_\rL v^i$ and $v^{\bar i} \sim c_\rR v^{\bar i}$, and therefore 
they parametrise  identical complex projective 
spaces $ {\mathbb C}P^1_\rL $ and ${\mathbb C}P^1_\rR$.
The superfields $G_\rL^{(2)}(v_\rL)$ and $G_\rR^{(2)}(v_\rR)$
are naturally defined on curved $\cN=4$ projective superspace 
$\cM^{3|8} \times {\mathbb C}P^1_\rL \times {\mathbb C}P^1_\rR$ 
introduced in \cite{KLT-M11}.

The field strengths $G_\rL^{(2)}(v_\rL)$ and $G_\rR^{(2)}(v_\rR)$ are examples 
of the covariant projective multiplets introduced in \cite{KLT-M11} in SO(4) superspace
and later reformulated  in \cite{KN14} within the conformal superspace setting. 
There are two types of covariant projective multiplets, 
the left and right ones. A left projective multiplet of weight $n$, $Q^{(n)}_\rL (v_\rL)$,
is a superfield that is defined on some open domain of $ {\mathbb C}^2 \setminus  \{0\}$
and possesses the following four properties. Firstly, it is a primary superfield, 
\be S_\a^{i \bar{i}} Q_{\rm L}^{(n)} = 0 \ , \quad K_a Q_{\rm L}^{(n)} = 0 \ .
\ee
Secondly, it is subject to the constraint 
\bea
\nabla_\a^{(1)\bar i} Q^{(n)}_\rL =0 \ .
\label{220}
\eea
Thirdly, it is a {\it holomorphic} function of $v_\rL$.  
Fourthly, it is 
{\it homogeneous} function 
of $v_\rL$ of degree $n$, 
\be
Q_\rL^{(n)}(c\,v_\rL)\,=\,c^n\,Q_\rL^{(n)}(v_\rL)\ ,
\qquad c \in   {\mathbb C} \setminus  \{0\} \ .
\ee 
Every left projective multiplet is inert with respect to $\rm SU(2)_{\rm R}$ 
and transforms under $\rm SU(2)_{\rm L}$ as
\bsubeq
\bea
\d_\L Q_{\rm L}^{(n)} &=& \L^{ij} { L}_{ij} Q_{\rm L}^{(n)} \ , \\
\L^{ij}{ L}_{ij} Q_{\rm L}^{(n)} &=& - (\L^{(2)}_{\rm L} 
 \partial_{\rm L}^{(-2)} - n \L_{\rm L}^{(0)}) Q_{\rm L}^{(n)} \ ,
\eea
\esubeq
where we have defined
\be \L_{\rm L}^{(2)} := \L^{ij} v_i v_j \ , \quad \L_{\rm L}^{(0)}: = \frac{v_i u_j}{(v_{\rm L} , u_{\rm L})} \L^{ij}
\ee
and  made use of the differential operator
\be 
 \partial_{\rm L}^{(-2)} := \frac{1}{(v_{\rm L} , u_{\rm L})} u^i \frac{\partial}{\partial v^i} \ , \quad (v_{\rm L} , u_{\rm L}) = v^i u_i \ .
\ee
Here we have also introduced a second left isospinor variable $u_\rL:=u^{i}$ 
which is restricted to be linearly independent of $v_\rL$,
that is
$(v_\rL,u_\rL)\ne0$. 
One may see that $L^{(2)} Q_\rL^{(n)}=0$, and therefore the integrability condition 
$\big\{ \nabla_\a^{(1)\bar i} , \nabla_\b^{(1)\bar j} \big\} Q^{(n)} =0$ 
for the constraint \eqref{220}
holds, in accordance with 
\eqref{217}.
The right projective multiplets are defined similarly. 
The covariant projective multiplets $G_\rL^{(2)}(v_\rL)$ and $G_\rR^{(2)}(v_\rR)$ 
are known as the left and right
$\cO(2)$ multiplets, respectively.

As shown in \cite{KLT-M11} the self-dual vector multiplet, $G_{\rm R}^{(2)} (v_\rR) $, 
can be described 
in terms of a gauge prepotential $\cV_\rL(v_\rL)$, which is a left weight-0
{\it tropical multiplet} and is real with respect to the analyticity preserving 
conjugation called the smile conjugation. The interested reader is referred to \cite{KLT-M11} 
for the technical details.  
Similar properties hold 
for the anti-self-dual vector multiplet 
except all `left' objects have to be replaced by `right' ones and vice versa.

\subsection{Dynamics}

General off-shell matter couplings in $\cN=4$ supergravity were constructed in 
\cite{KLT-M11}.  
The action for such a supergravity-matter system
may be represented as a sum of two terms (one of which may be absent), 
\bea
S= S_{\rm L} + S_{\rm R}~.
\eea 
The left  $S_{\rm L}$ and right  $S_{\rm R}$ actions, are naturally 
formulated in  curved $\cN=4$ projective superspace. 
The left action has the form 
\bea
S_{\rL}  &=& \frac{1}{2\pi} \oint (v_\rL, \rd v_\rL)
\int 
\rd^{3|8} z
\,E\, C_\rL^{({-4})} \cL_\rL^{(2)}~, 
\qquad E^{-1}= {\rm Ber}(E_A{}^M)~,
\label{Action-left} 
\eea
where the Lagrangian $\cL_\rL^{(2)}(v_\rL)$ is a real left projective multiplet 
of weight 2, and $\rd^{3|8} z$ denotes 
the full superspace integration measure, 
$\rd^{3|8} z:= \rd^3 x \,{\rm d}^8\q$.
Furthermore, the {\it model-independent} primary
 isotwistor superfield  $C_\rL^{(-4)} (v_\rL) $ has dimension $-2$, i.e. 
 ${\mathbb D} C_\rL^{(-4)} =-2 C_\rL^{(-4)}$. 
It is defined to be real with respect to the smile-conjugation 
defined in \cite{KLT-M11} and obeys the differential equation
\bea
\D_\rL^{(4)}C_\rL^{(-4)}=1~.
\label{N=4AcComp-L}
\eea
Here $\D_\rL^{(4)}$ denotes the following fourth-order operator\footnote{The operator
$\D_\rL^{(4)}$ is a covariant projection operator. Given  a covariant left projective multiplet $Q_\rL^{(n)} (v_\rL) $ of weight $n$, 
it may be represented in the form 
$Q_\rL^{(n)}=\D_\rL^{(4)} T_\rL^{(n-4)}$, 
for some left isotwistor superfield $T_\rL^{(n-4)} (v_\rL )$, see  \cite{KLT-M11}
for details.}
\bea
\D_\rL^{(4)}&=&\frac{1}{96}\Big(
\nabla^{ (2) \bar i \bar j}  \nabla^{(2)}_{\bar i\bar j}
-\nabla^{(2)\a\b}  \nabla^{(2)}_{\a\b}
\Big) = \frac{1}{48}\nabla^{ (2) \bar i \bar j}  \nabla^{(2)}_{\bar i \bar j}
~,
\eea
with
$\nabla^{(2)}_{\bar i\bar j}:=\nabla^{(1)\g}_{(\bar i}\nabla^{(1)}_{\g \bar j)}$
and $\nabla^{(2)}_{\a\b}:=\nabla^{(1)\bar k}_{(\a}\nabla^{(1)}_{\b) \bar k}$.
The action \eqref{Action-left} is independent of the representative $ C_\rL^{(-4)} $ in the sense that 
it does not change under an arbitrary infinitesimal variation of  $C_\rL^{(-4)} $ subject to the above conditions. The structure of $S_\rR$ is analogous.

There are two equivalent action functionals to describe the dynamics of a single
self-dual Abelian vector multiplet coupled to conformal supergravity.
One of them is a right action formulated in terms of a right $\cO(2)$
multiplet $G_{\rm R}^{(2)}(v_{\rm R}) = v_{\bar i} v_{\bar j} G^{\bar{i} \,\bar{j}} $, 
which is associated with the 
superfield strength $G^{\bar{i} \,\bar{j}} $
of the vector multiplet. This action, has the form\footnote{We should emphasise that in this paper we have defined the vector multiplet actions with ``wrong'' 
sign,  because in our approach they correspond to superconformal compensators.} \cite{KLT-M11}
\bea
 S^{(+)}_{{\rm VM}} :=  \frac{\sqrt{2}}{\pi} \oint (v_{\rm R} , \rd v_{\rm R}) \int 
 \rd^{3|8} z
 \, E\,C_{\rm R}^{(-4)}
G^{(2)}_\rR \ln \frac{G^{(2)}_\rR }{\ri \Upsilon^{(1)}_\rR \breve\Upsilon^{(1)}_\rR } ~,
\label{actionplus}
\eea
where the weight-one arctic multiplet 
$\Upsilon^{(1)}_\rR $ and its smile conjugate
 $\breve\Upsilon^{(1)}_\rR $  are pure gauge degrees of freedom.  
The action \eqref{actionplus} is the 3D $\cN=4$ counterpart 
of the projective-superspace action \cite{K08} 
for the 4D $\cN=2$ improved tensor multiplet \cite{deWPV}.
The other representation for $S^{(+)}_{{\rm VM}} $ makes use of 
a left tropical prepotential 
$\cV_{\rm L}(v_{\rm L}) $ of the self-dual vector multiplet with gauge transformations 
\bea
\d  \cV_\rL = \l_\rL + \breve{\l}_\rL~.
\label{left-gauge}
\eea
The gauge parameter $\l_\rL$ is an arbitrary left arctic multiplet of weight zero.
The gauge invariant field strength, $G^{\bar i\bar j}$,
 is  related to $\cV_\rL$ through
\bea 
G_{\rm R}^{(2)}(v_{\rm R}) = v_{\bar i} v_{\bar j} G^{\bar{i} \,\bar{j}} 
= \frac{\ri}{4}  v_{\bar i} v_{\bar j}
\oint \frac{(v_{\rm L} , \rd v_{\rm L})}{2 \pi} 
\frac{u_iu_j}{(v_{\rm L}, u_{\rm L})^2}
 \nabla^{\a i \bar i} \nabla_\a{}^{j\bar j}
\cV_{\rm L}(v_{\rm L}) \ .
\label{D.10}
\eea
Here $u_\rL = u^i$ is a constant isospinor such that $(v_{\rm L}, u_{\rm L}) \neq 0$
along the closed integration contour.\footnote{One may show that the right-hand side of 
\eqref{D.10} is independent of $u_\rL$.}
The action \eqref{actionplus} can be recast as a left $BF$-type action \cite{KN14}
\be  \label{vm+}
S^{(+)}_{{\rm VM}} = -\frac{1}{2 \pi} \oint (v_{\rm L} , \rd v_{\rm L}) 
\int \rd^{3|8} z
\, E\,C_{\rm L}^{(-4)} \cV_{\rm L} \bm G_{\rm \rm L}^{(2)} \ ,
\ee
where 
 ${\bm G}_{\rm L}^{(2)}(v_{\rm L}) 
= v_{ i} v_{ j} {\bm G}^{{i} {j}} $ is the composite left
$\cO(2)$ multiplet \cite{KN14}
\bea
\bm G{}^{(2)}_{\rm L} &=& - \frac{\ri}{\sqrt{2}} v_{i} v_{j}
\oint \frac{(v_{\rm R} , \rd v_{\rm R})}{2 \pi} 
\frac{u_{\bar{i}} u_{\bar{j}}}{(v_{\rm R}, u_{\rm R})^2}
 \nabla^{\a i \bar i} \nabla_\a{}^{j\bar j}
\ln \frac{G^{(2)}_\rR }{\ri \Upsilon^{(1)}_\rR \breve\Upsilon^{(1)}_\rR } \non\\
&=& \frac{\ri}{4} v_{i} v_{j} \nabla^{\a i \bar{i}} \nabla_\a^{j \bar{j}} \Big( \frac{G_{\bar{i}\bar{j}}}{G_+} \Big) 
\ .
\eea
The  composite left superfield ${\bm G}^{{i} {j}} $ 
can be equivalently realised as the anti-self-dual SO(4) bivector 
${\bm G}_-^{IJ}$.

Similarly, the action for the anti-self-dual vector multiplet \cite{KLT-M11}
 can be recast as the right $BF$-type action \cite{KN14}
\be \label{vm-}
S_{\rm VM}^{(-)} :=- \frac{1}{2 \pi} \oint (v_{\rm R} , \rd v_{\rm R}) \int 
\rd^{3|8} z
\,E\, 
 C_{\rm R}^{(-4)} \cV_{\rm R} \bm G_{\rm R}^{(2)} \ ,
\ee
where  ${\bm G}_{\rm R}^{(2)}(v_{\rm R}) 
= v_{\bar i} v_{\bar j} {\bm G}^{\bar{i} \,\bar{j}} $ is the composite right
$\cO(2)$ multiplet \cite{KN14}
\bea
\bm G_\rR^{(2)} 
&=& - \frac{\ri}{\sqrt{2}} v_{\bar i} v_{\bar j}
\oint \frac{(v_{\rm L} , \rd v_{\rm L})}{2 \pi} 
\frac{u_iu_j}{(v_{\rm L}, u_{\rm L})^2}
 \nabla^{\a i \bar i} \nabla_\a{}^{j\bar j} \ln \frac{G_\rL^{(2)}}{\ri \U_\rL^{(1)} \breve{\U}_{\rL}^{(1)}} \non\\
 &=& v_{\bar i} v_{\bar j} \frac{\ri}{4} \nabla^{\a i \bar{i}} \nabla_\a^{j \bar{j}} \Big( \frac{G_{ij}}{G_-}\Big)
\ ,
\label{223}
\eea
and $\cV_\rR(v_\rR) $ is the tropical prepotential 
 of the anti-self-dual vector multiplet.  
The  composite right superfield \eqref{223} 
can be equivalently realised as the self-dual SO(4) bivector 
$\bm G_+^{IJ}$.

The composite $\cO(2)$ multiplets can be expressed in terms of $SO(4)$ vector 
indices as follows \cite{KN14}
\bea 
\bm G_{\pm}^{IJ} = X_\mp^{IJ} \pm \hf \eps^{IJKL} X_{\mp KL} \ , \quad
 \hf \eps_{IJKL} \bm G_{\pm}^{KL}  = \pm\bm G_{\pm IJ} 
 \ , 
\eea
where we have defined
\begin{align} X_{\pm}^{IJ} &:= \frac{\ri}{6 G_{\pm}} \nabla^{\g [I} \nabla_{\g K} G_{\pm}^{J] K}
+ \frac{2 \ri}{9 G_{\pm}^3} \nabla^{\a P} G_{\pm KP} \nabla_{\a Q} G_\pm^{Q [I} G_\pm^{J] K} \ .
\end{align}
To show that $\bm G_{\pm}^{IJ}$ is primary and satisfies the Bianchi identity, 
the following identities prove useful
\bsubeq
\begin{align} G_{\pm}^{IK} G_{\pm JK} &= \hf \d^I_J G_{\pm}^2 \ , \label{inverseGpm} \\
\eps^{IJKL} G_{\pm LP} &= \mp 3 \d^{[I}_P G_{\pm}^{JK]} \ .
\end{align}
\esubeq

It is worth mentioning that the
two $\cN = 4$ linear multiplet actions (\ref{vm+})  and (\ref{vm-}) 
 are universal \cite{KN14} in the sense that all
known off-shell supergravity-matter systems (with the exception of pure conformal
supergravity) may be described using such actions with appropriately engineered composite $\cO(2)$ 
multiplets $\bm G_{\rm L}^{(2)}$ and $\bm G_{\rm R}^{(2)}$.


\section{Minimal topologically massive supergravity}

In this section we 
present two new supergravity-matter systems 
as models for
minimal topologically massive supergravity. 



\subsection{Action principle and equations of motion}

Our models for minimal topologically massive supergravity 
are described by 
$\cN = 4$ conformal supergravity coupled to a vector multiplet, 
either self-dual or anti-self-dual,  
via the following supergravity-matter actions:
\be
\label{SUGRA-matteractionN=4} 
\kappa S_{\pm} := \frac{1}{\mu} S_{\rm CSG} + S_{{\rm VM}}^{(\pm)} \ , 
\qquad \k^2=1~,
\ee
where $S_{\rm CSG}$ denotes the conformal supergravity 
action \cite{BKNT-M2}. We will refer to the theories with actions 
$S_+$ and $S_-$ as the self-dual and anti-self-dual topologically massive 
supergravity (TMSG) theories, respectively.

As shown in \cite{KN14}, the equation of motion for the vector multiplet 
derived from the action \eqref{SUGRA-matteractionN=4} is equivalent to
\be \bm G_{\mp}{}^{IJ} = 0 \ , \label{N=4EoM}
\ee
while the equation of motion for the conformal supergravity multiplet 
(that is, the $\cN=4$ Weyl supermultiplet) is
\be \label{SUGRAEoM}
\frac{1}{\mu} W + T_{\pm} = 0 \ .
\ee
Here $T_{\pm}$ is the supercurrent, which corresponds to the action 
$S_{{\rm VM}}^{(\pm)} $, 
\be \label{tg}
T_{\pm} = \pm G_{\pm} \ .
\ee
One can check that the supercurrent $T_{\pm}$ obeys the conservation equation \cite{KNT-M13}
\bea
\nabla^{\a (I} \nabla_\a^{J)} T_{\pm} = \frac{1}{4} \d^{IJ} \nabla^{\a}_K \nabla_\a^{K} T
\label{N=4scce}
\eea
when the matter equation of motion \eqref{N=4EoM} is satisfied.

Making use of the Bianchi identity \eqref{VMBI} as well as the equations of motion (\ref{N=4EoM})--(\ref{tg}), one 
finds the following equations on $G_{\pm}$:
\bsubeq \label{CSeqMonG}
\bea
\Big( \nabla^{\g (I} \nabla_{\g}^{J)} - \frac{1}{4} \d^{IJ} \nabla^\g_K \nabla_\g^K \Big) G_{\pm} &=& 0 \ , \\
\Big(\nabla^\g_K \nabla_\g^K \mp 8 \ri W \Big) G_{\pm}^{-1} &=& 0 \ , \\
\frac{1}{\mu} W \pm G_{\pm} &=& 0 \ , \\
\nabla_{(\a}^{[I} \nabla_{\b)}^{J]}  G_{\pm}^{-1}
&=& \pm \hf \eps^{IJKL} \nabla_{(\a K} \nabla_{\b) L}  G_{\pm}^{-1} \ .
\eea
\esubeq
We now turn to an analysis of 
the consequences of the equations of motion \eqref{CSeqMonG}.


\subsection{Analysing the equations of motion}

To analyse the equations of motion corresponding to the action \eqref{SUGRA-matteractionN=4} 
we need to fix the gauge freedom. Firstly, we use the special conformal 
transformations to make the dilatation connection vanish, $B_A = 0$. This corresponds to degauging of conformal superspace to SO(4) superspace \cite{KLT-M11} and gives rise to new torsion terms\footnote{See \cite{BKNT-M2} for more details.
It is important to note that the SO(4) connection of SO(4) superspace differs 
from the one of conformal superspace by a redefinition, for details see \cite{BKNT-M1}.} 
which can be expressed in terms of superfields $\cS^{IJ}$, $\cS$, $C_a{}^{IJ}$ and their covariant derivatives. We refer the reader 
to \cite{KLT-M11} for details and provide a summary of the salient details 
of SO(4) superspace in Appendix \ref{geometrySO4}.

Upon imposing the gauge $B_A = 0$ one can show that \eqref{CSeqMonG} is equivalent to
\bsubeq \label{SO(4)eqMonG}
\bea
\Big( \cD^{\g (I} \cD_{\g}^{J)} - \frac{1}{4} \d^{IJ} \cD^\g_K \cD_\g^K - 4 \ri \cS^{IJ} \Big) G_{\pm} &=& 0 \ , \\
\Big(\cD^\g_K \cD_\g^K + 8 \ri (2 \cS \mp W)\Big) G_{\pm}^{-1} &=& 0 \ , \\
\frac{1}{\mu} W \pm G_{\pm} &=& 0 \ , \\
(\cD_{(\a}^{[I} \cD_{\b)}^{J]} - 4 \ri C_{\a\b}{}^{IJ}) G_{\pm}^{-1}
= \pm \hf \eps^{IJKL} (\cD_{(\a K} \cD_{\b) L} &-& 4 \ri C_{\a\b KL}) G_{\pm}^{-1} \ ,
\eea
\esubeq
where $\cD_\a^I$ is the $\rm SO(4)$ superspace covariant derivative 
\cite{KLT-M11,HIPT} (see also \cite{BKNT-M1}). In isospinor index notation, 
for the self-dual vector multiplet one obtains
\bsubeq \label{isospinEoMS+}
\bea
\Big( \cD^{\g i\bar{i}} \cD_{\g i\bar{i}} + 8 \ri (2 \cS - W) \Big) G_{+}^{-1} &=& 0 \ , 
\\
(\cD_\a^{(\bar{i} \bar{k}} \cD_\b{}^{j)}{}_{\bar{k}} - 4 \ri C_{\a\b}{}^{ij} ) G_{+}^{-1} &=& 0 \ , \\
(\cD^{\g(i(\bar{i}} \cD_\g^{j)\bar{j})} - 4 \ri \cS^{ij\bar{i}\bar{j}}) G_{+} &=& 0 \ , 
 \\
W 
+\mu G_{+} &=&0\ ,
\eea
\esubeq
 while for the anti-self-dual vector multiplet one finds
\bsubeq
\bea
\Big( \cD^{\g i\bar{i}} \cD_{\g i\bar{i}} + 8 \ri (2 \cS + W) \Big) G_{-}^{-1} &=& 0 \ , 
\\
(\cD_\a^{k(\bar{i}} \cD_\b{}_{k}{}^{\bar{j})} - 4 \ri C_{\a\b}{}^{\bar{i}\bar{j}} ) G_{-}^{-1} &=& 0 \ , \\
(\cD^{\g(i(\bar{i}} \cD_\g^{j)\bar{j})} - 4 \ri \cS^{ij\bar{i}\bar{j}}) G_{-} &=& 0 \ ,
 \\
W 
-\mu G_{-} &=&0\ .
\eea
\esubeq
One should keep in mind that the equations of motion for $G_+$ and $G_-$ 
derived from the actions $S_+$ 
and $S_-$, respectively, were used in the above results.

Under super-Weyl transformations the $SO(4)$-covariant derivatives and the torsion terms transform as\footnote{The infinitesimal form was given in \cite{KLT-M11, KLT-M12}.}
\bsubeq
\bea
\cD_\a^I \rightarrow \cD'{}_\a^I &=&
\re^{\hf \s}\Big( \cD_\a^I + (\cD^{\b I}\s) M_{\a\b}+(\cD_{\a J} \s)N^{IJ} \Big)
~, \\
\cS^{IJ} \rightarrow \cS'^{IJ} 
&=& \frac{\ri}{4} \re^{2 \s} (\cD^{\g (I} \cD_{\g}^{J)} - \frac{1}{4} \d^{IJ} \cD^{\g K} \cD_{\g K} - 4 \ri \cS^{IJ}) \re^{- \s} 
~,
\\
\cS \rightarrow \cS' &=&
- \frac{\ri}{16} (\cD^\g_K \cD_\g^K + 16 \ri \cS) \re^\s
~, \\
C'_{a}{}^{IJ} \rightarrow C_a{}^{IJ} &=&
- \frac{\ri}{8} (\g_a)^{\a\b} (\cD_{\a}^{[I} \cD_\b^{J]} - 4 \ri C_{\a\b}{}^{IJ}) \re^\s
~, \\
W \rightarrow W' &=& \re^\s W \ ,
\eea
\esubeq
where $\s$ is a real unconstrained superfield. 
Within  the superconformal framework, all supergravity-matter actions are required to be super-Weyl invariant.

The super-Weyl  gauge freedom may be used to impose useful gauge conditions. 
For instance, one can make use 
of the super-Weyl transformations to gauge away  the self-dual 
or anti-self-dual part of $C_{a}{}^{IJ}$ such 
that the remaining torsion components are expressed directly in terms of the matter fields. 
For concreteness, let us consider the theory described by the action $S_{+}$, 
with corresponding equations of motion \eqref{isospinEoMS+},  
and gauge away $C_a{}^{\bar{i}\bar{j}}$ via a super-Weyl transformation.
We then find
\bsubeq
\bea
W &=& - \mu G_{+} \ , \\
\cS^{ij\bar{i}\bar{j}} &=& - \frac{\ri}{4} G_{+}^{-1} \cD^{\g (i(\bar{i}} \cD_\g^{j)\bar{j})} G_+ \ , \\
2 \cS - W &=& \frac{\ri}{8} G_+ \cD^{\g i \bar{i}} \cD_{\g i\bar{i}} G_+^{-1} \ , \\
C_{\a\b}{}^{ij} &=& - \frac{\ri}{4} G_+ \cD_\a^{(i \bar{k}} \cD_\b{}^{j)}{}_{\bar{k}} G_+^{-1} \ , \\
C_{a}{}^{\bar{i}\bar{j}} &=& 0 \ .
\eea
\esubeq
In this gauge, we see that the geometry is determined in terms of a single superfield, 
which is chosen to be the  scalar $G_+$. 
After imposing this super-Weyl gauge condition it is possible to show that there is enough super-Weyl freedom 
left to impose the additional condition 
\be 
2 \cS + W = 0~, 
\ee
see Appendix \ref{2VMtoCSG} for the derivation.
This condition proves to lead to the following 
nonlinear equation for $G_+$:
\bea 
\cD^{\g i \bar{i}} \cD_{\g i \bar{i}} G_+^{-1} + 16 \ri \mu = 0 \ . 
\label{G+Equation}
\eea

The main virtue of the super-Weyl gauge conditions imposed is that 
all the torsion and curvature tensors are descendants of the single scalar superfield
$G_+$. However, this gauge choice is not particularly useful from the point of view 
of studying (maximally) supersymmetric backgrounds. A more convenient super-Weyl gauge fixing is  $G_+ = {\rm const}$. 
We spell out the implications of such a gauge condition below.

Given a vector multiplet with a superfield strength $G^{IJ}$ such that $G$ is nowhere 
vanishing, one can always make use of the 
super-Weyl transformations to choose a gauge where
\be
G = \hf G^{IJ} G_{IJ} = 1 \ , \quad \cD_\a^I G^{JK} = 0 \ .
\ee
Such a gauge condition has slightly different consequences on the superspace geometry for the two vector multiplets 
$G_+^{IJ}$ and $G_{-}^{IJ}$ satisfying the equations of motion \eqref{N=4EoM} and \eqref{SUGRAEoM}. 
In both cases the super-Cotton tensor is constant,
\be W = {\rm const} \quad \Longrightarrow \quad \cS^{IJ} = 0 \ ,
\ee
while the constraints on the remaining torsion components differ. For the on-shell self-dual vector multiplet one finds 
the following consistency conditions
\be 
\hf \eps_{IJKL} C_{a}{}^{KL} = C_{a}{}_{IJ} \ ,  \quad 
2 \cS - W = 0 \ ,
\ee
while for the on-shell anti-self-dual vector multiplet one finds
\be 
- \hf \eps_{IJKL} C_{a}{}^{KL} = C_{a}{}_{IJ} \ , \quad 
2 \cS + W = 0 \ .
\ee
In the case where $C_{a}^{IJ}$ vanishes, the algebra of covariant derivatives coincides with that of 
$(4, 0)$ AdS superspace in the critical case where $2 \cS \mp W = 0$, see \cite{KLT-M12}.\footnote{The $\cN=4$ super-Cotton 
tensor is denoted by $X$ in \cite{KLT-M11,KLT-M12} .} 
In general, however, $C_a{}^{IJ}$ does not vanish and instead 
satisfies some differential conditions implied by the Bianchi identities
\bea \label{covDerBI}
[[\cD_A ,\cD_B\} , \cD_C \} &+& (-1)^{\eps_A (\eps_B + \eps_C)} [[\cD_B ,\cD_C\} , \cD_A \} \non\\
&+& (-1)^{\eps_C (\eps_A + \eps_B)} [[\cD_C ,\cD_A\} , \cD_B \} = 0 \ .
\eea
To analyse the Bianchi identities in detail it will be useful to convert to isospinor notation.

We consider in detail the self-dual TMSG theory. 
In the isospinor notation, the covariant derivative algebra 
which follows from the equations of motion is
\bsubeq 
\label{CovDerAlgGaugeFixed}
\bea
\{ \cD_\a^{i\bar{i}} , \cD_\b^{j\bar{j}}\}
&=&
2 \ri \eps^{ij} \eps^{\bar{i}\bar{j}} \cD_{\a\b}
+ 4 \ri \eps_{\a\b} \eps^{\bar{i}\bar{j}} W L^{ij}
+ 4 \ri C_{\a\b}{}^{\bar{i}\bar{j}} L^{ij} \non\\
&&+ 2 \ri \eps_{\a\b} \eps^{ij} C^{\g\d \bar{i}\bar{j}} M_{\g\d}
- 2 \ri \eps^{ij} \eps^{\bar{i} \bar{j}} W M_{\a\b} \ .
\eea
Analysing the Bianchi identities \eqref{covDerBI} 
determines the remainder of the covariant derivative algebra:
\bea
\left[ 
\cD_{\a\b} , \cD_\g^{k \bar{k}}
\right]
&=&
- \eps_{\g(\a} W \cD_{\b)}^{k \bar{k}}
+ (\eps_{\g(\a} C_{\b)\d}{}^{\bar{k}\bar{j}} + \eps_{\d(\a} C_{\b)\g}{}^{\bar{k}\bar{j}}) \cD^\d{}^k{}_{\bar{j}}
\non\\
&&+ 2 \eps_{\g(\a} C_{\b)\d\r}{}^{k\bar{k}} M^{\d\r}
- 2 C_{\a\b\g}{}^{j\bar{k}} L_j{}^k \ , \\
\left[ \cD_{\a\b} , \cD_{\g\d} \right] &=&
\ri \eps_{\g(\a} C_{\b)\d\r}{}_{k \bar{k}} \cD^{\r k \bar{k}}
+ \ri \eps_{\d(\a} C_{\b)\g\r}{}_{k\bar{k}} \cD^\r{}^{k\bar{k}} \non\\
&&+ \eps_{\d(\a} W^2 M_{\b)\g} +  \eps_{\g(\a} W^2 M_{\b)\d}
\non\\
&&
+ \frac{\ri}{12} \eps_{\d(\a} \Big( \cD_{\b)}^{k\bar{k}} \cD_{\g}{}_k{}^{\bar{l}} C_{\r\s}{}_{\bar{k}\bar{l}} \Big) M^{\r\s}
+ \frac{\ri}{12} \eps_{\g(\a} \Big( \cD_{\b)}^{k\bar{k}} \cD_{\d}{}_k{}^{\bar{l}} C_{\r\s}{}_{\bar{k}\bar{l}} \Big) M^{\r\s}
\non\\
&&- \eps_{\d(\a} C_{\b)\g \bar{k}\bar{l}} C^{\r\s \bar{k}\bar{l}} M_{\r\s}
- \eps_{\g(\a} C_{\b)\d \bar{k}\bar{l}} C^{\r\s \bar{k}\bar{l}} M_{\r\s} \ ,
\eea
\esubeq
as well as the following differential constraint on $C_{a}{}^{\bar{i}\bar{j}}$
\be \label{BIonCSO(4)}
\cD_\a^{i \bar{i}} C_{\b\g}{}^{\bar{j}\bar{k}}
= 2 \eps^{\bar{i} (\bar{j}} C_{\a\b\g}{}^{i \bar{k})} \ .
\ee
The above constraint implies, in turn,
\be \label{massiveCequation}
\cD_\a{}^{\g} C_{\b\g}{}^{\bar{i}\bar{j}} + C_{(\a}{}^{\g}{}_{\bar{k}}{}^{(\bar{i}} C_{\b) \g}{}^{\bar{j})\bar{k}}
+ 2 W C_{\a\b}{}^{\bar{i}\bar{j}} = 0 \ .
\ee
Since the SU$(2)_{\rm R}$ curvature vanishes, we can completely gauge away 
the corresponding connection. Such a gauge condition is assumed in what follows.
In this gauge, the field strength  $G^{\bar i \bar j}$ becomes a constant
symmetric isospinor subject to the normalisation condition
 $G^{\bar i \bar j}G_{\bar i \bar j}=1$. It is invariant under a U(1) 
 subgroup of SU$(2)_{\rm R}$. 

We are now in a position to describe 
all maximally supersymmetric solutions of the theory. 
In accordance with the general superspace analysis of supersymmetric backgrounds in diverse dimensions \cite{K13,KNT-M,K15Corfu}, 
such superspaces have to comply with the additional constraint
\bea
\cD_\a^{i \bar{i}} C_{\b\g}{}^{\bar{j}\bar{k}} 
=0~, 
\label{322}
\eea
which leads to the integrability conditions
\begin{subequations}
\bea
(\cD_a -W M_a) C_b{}^{\bar{j}\bar{k}} &=&0~, \label{323a} \\
C^\g{}_{(\a}{}^{\bar{i}\bar{j}} C_{\b)\g}{}^{\bar{k}\bar{l}}&=&0~. \label{323b}
\eea
\end{subequations}
The general solution of \eqref{323b} is
\bea
C_{\a\b}{}^{\bar{i}\bar{j}} = C_{\a\b} C^{\bar i \bar j}~, 
\eea
where $C^{\bar i \bar j}$ is a constant symmetric rank-2 isospinor.
Without loss of generality, $C^{\bar i \bar j}$ can be normalised as
$C^{\bar i \bar j}C_{\bar i \bar j}=1$.  The covariant constancy conditions \eqref{322} and \eqref{323a}
now amount to 
\bea
 \cD_\a^{i \bar{i}} C_b =0~, \quad (\cD_a -W M_a) C_b=0~.
 \label{3.25}
\eea
We recall that the Lorentz generator with a vector index, $M_a$, acts on a three-vector 
by the rule $M_a C_b = \ve_{abc} C^c$. 
The second condition in \eqref{3.25} implies that $C_b$ is a Killing vector of constant norm, 
\bea
\cD_a C_b + \cD_b C_a =0 ~ ,\quad C^2 =C^a C_a ={\rm const}~.
\eea
Thus there are three types of backgrounds depending on whether
the Killing vector $C^a$ is chosen to be time-like, space-like or null. 
The algebra of covariant derivatives for such a background is
\bsubeq 
\label{327}
\begin{align}
\{ \cD_\a^{i\bar{i}} , \cD_\b^{j\bar{j}}\}
&=
2 \ri \eps^{ij} \eps^{\bar{i}\bar{j}} (\cD_{\a\b} -W M_{\a\b})
+ 4 \ri \eps_{\a\b} \eps^{\bar{i}\bar{j}} W L^{ij}
+ 4 \ri C^{\bar{i}\bar{j} }  C_{\a\b} L^{ij} 
\non\\
&+ 2 \ri \eps_{\a\b} \eps^{ij}  C^{ \bar{i}\bar{j}}C^{\g\d  } M_{\g\d}
\ , \\
\left[ 
\cD_{\a\b} , \cD_\g^{k \bar{k}}
\right]
&=
- \eps_{\g(\a} W \cD_{\b)}^{k \bar{k}}
+ (\eps_{\g(\a} C_{\b)\d}{}^{\bar{k}\bar{j}} + \eps_{\d(\a} C_{\b)\g}{}^{\bar{k}\bar{j}}) \cD^\d{}^k{}_{\bar{j}}
\ , \\
\left[ \cD_{\a\b} , \cD_{\g\d} \right] &=
W^2\big ( \eps_{\d(\a}  M_{\b)\g} +  \eps_{\g(\a}  M_{\b)\d}\big)
- \big( \eps_{\d(\a} C_{\b)\g } 
+ \eps_{\g(\a} C_{\b)\d } \big)C^{\r\s } M_{\r\s} \ .
\end{align}
\esubeq
One may think of this algebra as a Lie superalgebra.\footnote{More precisely, 
\eqref{327} is isomorphic to  the Lie superalgebra corresponding to the isometry supergroup of the background superspace under consideration.} 
By construction, the theory involves the constant symmetric isospinor 
$G^{\bar i \bar j}$ being invariant under a U(1) subgroup of the group SU$(2)_{\rm R}$. 
If $C^{\bar i \bar j}$ does not coincide with $G^{\bar i \bar j}$, then the group
SU$(2)_{\rm R}$ is completely broken. This indicates that 
$C^{\bar i \bar j}=G^{\bar i \bar j}$.

The simplest maximally supersymmetric solution of the theory
is characterised by (see also \cite{LS16}) 
\bea
C_{a}{}^{\bar{i}\bar{j}} = 0~.
\eea
It corresponds to the critical (4,0) AdS superspace
introduced in \cite{KLT-M12}. Its algebra of covariant derivatives is
as follows:
\bsubeq 
\bea
\{ \cD_\a^{i\bar{i}} , \cD_\b^{j\bar{j}}\}
&=&
2 \ri \eps^{ij} \eps^{\bar{i}\bar{j}} ( \cD_{\a\b} -W M_{\a\b})
+ 4 \ri \eps_{\a\b} \eps^{\bar{i}\bar{j}} W L^{ij}
\ ,\\
{[}\cD_a,\cD_\b^{j\baj}{]}&=& \hf W (\g_a)_\b{}^\g\cD_\g^{j\baj}~,~~~~~~
{[}\cD_a,\cD_b{]}= - W^2 M_{ab}~.
\eea
\end{subequations}
The last relation shows that the cosmological constant is $\L = - W^2 = -\ell^{-2}$, in agreement with \cite{LS16,KLT-M12}. Here $\ell$ is the radius of curvature in AdS${}_3$. The latter relation is equivalent to 
$\m \ell=1$, which corresponds to chiral gravity \cite{LSS}.

More generally, the $(p,q)$ AdS superspaces, $p+q = \cN$, in three dimensions 
were classified in \cite{KLT-M12}.\footnote{In three dimensions,
 $\cN$-extended AdS supergravity exists in several incarnations  \cite{AT}
 known as the  $(p,q)$ AdS supergravity theories,
where the integers $p \geq q \geq 0$ are such that $\cN=p+q$. } 
In the $\cN=4$ case, the (3,1) and (2,2) AdS superspaces are necessarily
conformally flat, $W=0$.   The distinguished feature of (4,0) AdS supersymmetry 
 is that the super-Cotton scalar $W $ may have a non-zero value. The algebra of covariant derivatives 
 is  given by \cite{KLT-M12}
\bsubeq \label{4200}
\bea
\{\cD_\a^{i\bai},\cD_\b^{j\baj }\}&=&
2\ri\ve^{ij}\ve^{\bai \baj }\cD_{\a\b}
+{2\ri}\ve_{\a\b}\ve^{\bai \baj }(2\cS+W)L^{ij}
+2\ri\ve_{\a\b}\ve^{ij}(2\cS-W)R^{\bai \baj }
\non\\
&&-4\ri  \cS\ve^{ij}\ve^{\bai \baj }M_{\a\b}~,
\\
{[}\cD_a,\cD_\b^{j\baj}{]}&=& \cS(\g_a)_\b{}^\g\cD_\g^{j\baj}~,~~~~~~
{[}\cD_a,\cD_b{]}= - 4\cS^2M_{ab}~,
\eea
\esubeq
where the positive constant $\cS$ determines the curvature of AdS${}_3$.
For a generic value of $W$ the 
entire SO(4) $R$-symmetry group 
belongs to the superspace holonomy group.
But there are two special values of $W$ for which either the SU(2)$_\rR$ or the SU(2)$_\rL$ curvature vanishes
and the structure group is reduced.
These are given by
\bea
W=\pm 2 \cS
\eea
and correspond to the  critical (4,0) AdS superspaces.
As briefly discussed in \cite{KT-M14}, 
the isometry group of (4,0) AdS superspace is isomorphic
to ${\rm D} (2, 1;\a) \times {\rm SL}(2,{\mathbb R})$ 
in the non-critical case  $W\neq \pm 2\cS$,
where ${\rm D} (2, 1;\a)$ is one of the exceptional simple supergroups, 
with the real number $\a \neq  -1, 0  $, see e.g. \cite{DeWitt,FSS} for reviews. 
The supergroup parameter $\a$ 
is related to the (4,0) AdS parameter $q=1 +\frac{W}{2\cS}$ introduced in \cite{KT-M14}.
If the values of $\a$ are restricted to the range\footnote{Not all values of $\a$ 
lead to distinct supergroups, since the supergroups defined by the parameters $\a^{\pm 1}$,
$-(1+\a)^{\pm 1} $ and $-\a^{\pm 1}(1+\a)^{\mp 1}$  
are isomorphic \cite{DeWitt,FSS}.}
$-1 < \a \leq -\hf$, then  we can identify $-2\a = 1 +\frac{W}{2\cS}$. 
The case $\a =-\hf$  corresponds to the conformally flat (4,0) 
AdS superspace, for which $W=0$. 
Its  isometry group is 
${\rm OSp}(4|2) \times {\rm SL}(2,{\mathbb R})$.
 The  limiting choice $\a =-1$  corresponds to one  of the two critical  (4,0) AdS cases,
 $W=2 \cS$.\footnote{The isometry groups of the two critical
(4,0) AdS superspaces are isomorphic.}
The isometry group of this (4,0) AdS superspace is 
${\rm SU} (1, 1|2)  \rtimes {\rm SU} (2) \times {\rm SL}(2,{\mathbb R})$, see also the discussion in \cite{BILS}.

If $C^a \neq 0$, the maximally supersymmetric background \eqref{327} 
describes a warped critical (4,0) AdS superspace. The bosonic body of such 
a superspace is warped AdS${}_3$ spacetime associated with the Killing vector 
$c^a (x) = C^a (z)|_{\q=0}$. Warped  AdS${}_3$ spacetimes have been discussed in 
detail in the literature, see \cite{APSS,DKSS2,KLL,DM,M} and references therein. 
In the $\cN=2$ supersymmetric case, 
the (super)space geometry of maximally supersymmetric warped 
(2,0) AdS backgrounds was described in \cite{KLRST-M} and further elaborated in \cite{K15Corfu}.  Supersymmetric warped 
(1,1) AdS backgrounds, which are necessarily non-maximal,
 were thoroughly studied  in \cite{DKSS2}. 
 
 It is worth giving a few general comments about
maximally supersymmetric
warped AdS backgrounds in $\cN$-extended supergravity theories.
Such backgrounds do not exist
in the case of $\cN=1$ supergravity. 
This result was first demonstrated by Gibbons, Pope and Sezgin \cite{GPS},  
and it follows trivially from the general superspace analysis 
of supersymmetric backgrounds in diverse dimensions 
\cite{K13,KNT-M,K15Corfu}.\footnote{Indeed, the superspace
geometry of $\cN=1$ supergravity is determined by two torsion tensors, 
a scalar $S$ and a symmetric spinor $C_{\a\b\g} =C_{(\a\b\g)}$, 
see \cite{GGRS,KLT-M11} for more details.
According to \cite{KNT-M,K15Corfu}, every maximally supersymmetric background
is characterised by the conditions $C_{\a\b\g}=0$ and $S={\rm const}$, 
see also  \cite{KNT-M15}.
The resulting algebra of covariant derivatives corresponds 
to $\cN=1$ AdS superspace for $S\neq 0$, or Minkowski superspace for $S=0$.}
However, 
maximally supersymmetric
warped AdS backgrounds do exist 
in extended supergravity, $\cN>1$, 
if the structure group includes not only the Lorentz group $ {\rm SL}(2,{\mathbb R})$
but also a nontrivial 
$R$-symmetry group. For instance, the structure group for 
 $\cN=(2,0)$ AdS supergravity is $ {\rm SL}(2,{\mathbb R}) \times {\rm U(1)_R}$, 
 and thus this theory
possesses  maximally supersymmetric warped  AdS backgrounds, which were described in \cite{KLRST-M,K15Corfu} using the superspace techniques, 
and some time later in \cite{KLL,DM} using the component approach. 
On the other hand,  the structure group for 
 $\cN=(1,1)$ AdS supergravity coincides with the Lorentz group, 
 and therefore this theory
possesses  no maximally supersymmetric warped  AdS backgrounds,
see \cite{KLRST-M,K15Corfu} for more details.

We now linearise the equation \eqref{massiveCequation} around the
critical  $(4, 0)$ AdS superspace and 
let $C_a{}^{\bar{i}\bar{j}} = \delta C_a{}^{\bar{i}\bar{j}}$ where $\delta C_a{}^{ij}$ is a small disturbance. Eq. \eqref{massiveCequation} 
turns into 
\be {\bm \cD}_\a{}^{\g} \delta C_{\b\g}{}^{\bar{i}\bar{j}} 
- 2 \mu \delta C_{\a\b}{}^{\bar{i}\bar{j}} = 0 \quad \Longrightarrow \quad
 {\bm \cD}^{a} \delta C_{a}{}^{\bar{i}\bar{j}} =0\ ,
 \label{332}
\ee
where ${\bm \cD}_a$ denotes the vector covariant derivative of the 
critical $(4, 0)$ AdS superspace. 
After applying another vector derivative one finds the equation
\be 
({\bm \cD}^a {\bm \cD}_a - 2 \mu^2 ) \delta C_{b}{}^{\bar{i}\bar{j}} = 0 \ .
\ee

One can also derive further equations 
on descendants of $\delta C_{\a\b}{}^{\bar{i}\bar{j}}$ using the 
constraint \eqref{BIonCSO(4)}. In particular, one finds
\bsubeq \label{CsmassiveEqs}
\bea
(\bm \cD_\a{}^\d - \frac{3}{2} \mu \d_\a^\d) \delta C_{\b\g\d}{}^{i \bar{i}} &=& 0 \ ,
\quad \delta C_{\a\b\g}{}^{i \bar i} := \frac{1}{3} \bm \cD_\a{}^i{}_{\bar{j}} \delta C_{\b\g}{}^{\bar{i}\bar{j}} \ ,
\label{334a} \\
(\bm \cD_\a{}^{\r} - \mu \d_\a^\r ) \delta C_{\b\g\d\r} &=& 0 \ , \quad
\delta C_{\a\b\g\d} := {\bm \cD}_{(\a}^{i\bar{i}} \delta C_{\b\g\d) i\bar{i}} \ ,
\label{334b}
\eea
\esubeq
where 
$\bm \cD_\a^{i\bar{i}}$ denotes the spinor covariant derivative of the 
critical $(4, 0)$ AdS superspace. The 
component projection of $\delta C_{\a\b\g}{}^{i\bar{i}}$ is proportional to the linearised gravitino field strength, 
while $\delta C_{\a\b\g\d}$ is proportional to the linearised Cotton tensor. These superfields can be 
shown to satisfy the following consequences of eqs. \eqref{CsmassiveEqs}:
\bsubeq
\bea
({\bm \cD}^a {\bm \cD}_a + \frac{1}{4} \mu^2) \delta C_{\a\b\g}{}^{i \bar{i}} &=& 0 \ , \\
({\bm \cD}^a {\bm \cD}_a + 2 \mu^2) \delta C_{\a\b\g\d} &=& 0 \ .
\eea
\esubeq
In the above we made use of the following result for a symmetric rank-$(2s)$ superfield $T_{\a_1 \cdots \a_{2s}} = T_{(\a_1 \cdots \a_{2s})}$ 
(with isospinor indices suppressed):
\be
(\bm \cD_{\a_1}{}^{\b} - \d_{\a_1}{}^\b \frac{\m}{\eta}) T_{\a_2 \cdots \a_{2s} \b} = 0 \quad 
\Longrightarrow \quad ({\bm \cD}^a {\bm \cD}_a - \frac{\m^2}{\eta^2} + (s+1) \mu^2) 
T_{\a_1 \cdots \a_{2s}} = 0
\ ,
\label{336}
\ee
with $\eta$ a dimensionless parameter. Computing the bar-projection of the 
equations \eqref{332}, \eqref{334a} and \eqref{334b}, we can determine 
the representations of the AdS group SO(2,2) to which 
the fields $ \delta C_{\a\b}{}^{\bar{i}\bar{j}} |$, $ \delta C_{\a\b\g}{}^{i \bar{i}}|$ and  
$ \delta C_{\a\b\g\d} |$ belong. We recall that the unitary representations of SO(2,2), 
denoted $D(E_0, \hat s)$,  
are labelled by two real weights $(E_0, \hat s)$, where $E_0$ is the lowest energy 
and $\hat s$ is the helicity, see e.g. \cite{DKSS}.  The weights obey the unitarity bound
$E_0 \geq |\hat s|$ for $\hat s >0$, where the representations 
with $E_0 =|\hat s| >0$ are called singleton representations. 
For a superfield $T_{\a_1 \dots \a_{2s}}$ obeying 
the first-order equation \eqref{336}, its lowest component $T_{\a_1 \dots \a_{2s}}|$ 
transforms in the representation with 

\bea
E_0 = 1 + \frac{1}{|\eta|} ~, \quad \hat s = \frac{s \eta }{|\eta|} ~,
\eea
as follows from the analysis in \cite{DKSS} (see also \cite{Bergshoeff:2010iy}).
Thus the gravitational field $ \delta C_{\a\b\g\d} |$  is a helicity 2 singleton, while 
the spin-1 and spin-3/2 fields,  
$ \delta C_{\a\b}{}^{\bar{i}\bar{j}} |$ and $ \delta C_{\a\b\g}{}^{i \bar{i}}|$, are massive.

In the above we worked with the self-dual TMSG theory, 
however the analysis of the equations of motion 
corresponding to the action $S_{-}$ is completely analogous. There one finds the covariant derivative algebra is
\bea
\{ \cD_\a^{i\bar{i}} , \cD_\b^{j\bar{j}}\}
&=&
2 \ri \eps^{ij} \eps^{\bar{i}\bar{j}} \cD_{\a\b}
- 4 \ri \eps_{\a\b} \eps^{ij} W R^{\bar{i}\bar{j}}
+ 4 \ri C_{\a\b}{}^{ij} R^{\bar{i}\bar{j}} \non\\
&&+ 2 \ri \eps_{\a\b} \eps^{\bar{i}\bar{j}} C^{\g\d ij} M_{\g\d}
+ 2 \ri \eps^{ij} \eps^{\bar{i} \bar{j}} W M_{\a\b} \ ,
\eea
where $C_a{}^{ij}$ satisfies the Bianchi identity
\be \cD_\a^{i\bar{i}} C_{\b\g}{}^{jk} = 2 \eps^{i(j} C_{\a\b\g}{}^{k)\bar{i}} \ .
\ee
Using the above equation one finds
\be \label{massiveBequation}
\cD_\a{}^{\g} C_{\b\g}{}^{ij} + C_{(\a}{}^{\g}{}_{k}{}^{(i} C_{\b) \g}{}^{j)k}
- 2 W C_{\a\b}{}^{ij} = 0 \ .
\ee
The solution $C_{a}{}^{ij} = 0$ corresponds to $(4, 0)$ AdS superspace 
in the critical case $2 S = - W$. We now linearise around the $(4, 0)$ AdS superspace and set $C_a{}^{ij} = \d C_a{}^{ij}$
where $\d C_a{}^{ij}$ is a small disturbance. It can be seen that $\d C_a{}^{ij}$ obeys the equation
\be 
{\bm \cD}_\a{}^{\g} \d C_{\b\g}{}^{ij} 
- 2 \mu \d C_{\a\b}{}^{ij} = 0 \ ,
\ee
where ${\bm \cD}_a$ corresponds to the vector covariant derivative of the $(4, 0)$ AdS superspace. After applying another 
vector derivative one finds
\be ({\bm \cD}^a {\bm \cD}_a - 2 \mu^2 ) \d C_{b}{}^{ij} = 0 \ .
\ee


\section{Component actions}

In this section we give the component results corresponding to the minimal 
$\cN = 4$ topologically massive 
supergravity action \eqref{SUGRA-matteractionN=4}.


\subsection{The component conformal supergravity action}

The complete component analysis of the $\cN$-extended Weyl multiplet
was given in \cite{BKNT-M2}. Here we specialise to the $\cN = 4$ case where the auxiliary fields 
coming from the super-Cotton tensor are defined as:
\bsubeq \label{defCompW}
\begin{align} w &:= \frac{1}{4!} \eps_{IJKL} w^{IJKL} = W| \ , \qquad
	y := \frac{1}{4!} \eps_{IJKL} y^{IJKL} = -\frac{\ri}{4} \nabla^\alpha_I \nabla_\alpha^I W |\ , \\
w_{\a L} &:= \frac{1}{3!} \eps_{IJKL} w_\a{}^{IJK} = - \frac{\ri}{2} \nabla_{\a L} W|~.
\end{align}
\esubeq

The full $\cN=4$ conformal supergravity action was given 
in \cite{BKNT-M2} and is
\begin{align} \label{N=4CSGActioN}
S_{\rm CSG} =& \  \frac{1}{8} \int \rd^3 x \,e \,\Big\{ \eps^{abc} \big( \omega_{a}{}^{fg} \cR_{bc}{}_{fg} - \frac{2}{3} \omega_{af}{}^g \omega_{bg}{}^h \omega_{ch}{}^f 
-  \frac{\ri}{2} \Psi_{bc}{}^\a_I (\g_d)_\a{}^\b (\g_a)_\b{}^\g \eps^{def} \Psi_{ef}{}^I_\g  \non\\
&- 2 {\cR}_{ab}{}^{IJ} V_c{}_{IJ} - \frac{4}{3} V_a{}^{IJ} V_b{}_I{}^K V_c{}_{KJ} \big) \non\\
&- 32 \ri w^\a_I w_\a^I - 8 w y
	- 16 \ri \,\psi_a{}^\a_I (\g^a)_{\a}{}^\b w_\b^{I} w - 2 \ri \eps^{abc} (\g_a)_{\a\b} \psi_b{}^\a_I \psi_c{}^{\b I} w^2 \Big\} \ ,
\end{align}
where the component curvatures $\cR_{ab}{}^{cd}$ and $\cR_{ab}{}^{IJ}$  are defined as 
\bsubeq
\begin{align}
\cR_{ab}{}^{cd} &:= 2 e_a{}^m e_b{}^n \partial_{[m} \omega_{n]}{}^{cd} - 2 \omega_{[a}{}^{cf} \omega_{b]}{}_f{}^d 
~,
\\
{\cR}_{ab}{}^{IJ} &:= 2 e_a{}^m e_b{}^n \partial_{[m} V_{n]}{}^{IJ} - 2 V_{[a}{}{}^{IK} V_{b] K}{}^J \ .
\end{align}
\esubeq


\subsection{The component vector multiplet actions} \label{CompVMactions}

The component $\cN=4$ linear multiplet actions were given in \cite{KN14}. Making use of 
the results there, one can construct the left and right vector multiplet actions.

The component fields of the vector multiplets are defined as
\bsubeq \label{N=4comps}
\begin{align}
g_\pm^{IJ} &:= G_\pm^{IJ}| \ , \\
\l_{(\pm)}{}_\a^I &:= \frac{2}{3} \nabla_{\a J} G_\pm^{IJ}| \ , \\
h_{(\pm)}{}^{IJ} &:= \frac{\ri}{3} \nabla^{\g [I} \nabla_{\g K} G_\pm^{J] K}| \ , \\
f_{(\pm) ab} &:=
- \frac{\ri}{24} \eps_{abc} (\g^c)^{\a\b} \nabla_\a^K \nabla_\b^L G_{\pm KL}|
- \hf (\psi_{[a}{}^K \g_{b]} \l_{(\pm) K})
+ \frac{\ri}{2} \psi_a{}^{\g K} \psi_b{}_\g^L \, g_{\pm KL} \ ,
\end{align}
\esubeq
where $g_{\pm}{}^{IJ}$ is (anti-)self-dual
\be \hf \eps^{IJKL} g_{\pm KL} = \pm g_{\pm}{}^{IJ} \ . \label{5.14}
\ee
The component gauge one-forms $v_{(\pm) a}$ are defined as
\be v_{(\pm) a} := e_a{}^m v_{(\pm) m} \ , \quad f_{(\pm) ab} = 2 e_a{}^m e_b^n \, \partial_{[m} v_{(\pm) n]} \ , \quad v_{(\pm) m} := V_{\pm m}| \ ,
\ee
where $V_{\pm}$ is the superspace gauge one-form associated with the field strength $G_{\pm}^{IJ}$.

It is useful to replace $h_{(\pm)}{}^{IJ}$ by the fields
\begin{align} \hat{h}_{\pm}{}^{IJ} &= \hf (h_{(\mp)}{}^{IJ} + \tilde{h}_{(\mp)}{}^{IJ}) \non\\
&= h_{(\mp)}{}^{IJ} \mp 2 w g_{\mp}{}^{IJ} \ ,
\end{align}
which proves to be (anti-)self-dual
\be \hf \eps^{IJKL} \hat{h}_{\pm KL} = \pm \hat{h}_{\pm}^{IJ} \ .
\ee

The component self-dual vector multiplet action is
\begin{align}
S_{\rm VM}^{(+)} =& - \int \rd^3 x \,e \Big( \eps^{abc} v_{(+) a} \bm f_{(+) bc} + \frac{1}{4} \hat{\bm h}_+{}^{IJ} g_{+ IJ} 
+ \frac{1}{4} \hat{h}_-{}^{IJ} \bm g_{-IJ} - \frac{\ri}{2} \l^{\a I} \bm \l_{\a I} \non\\
&\quad- \frac{1}{2} (\g^a)_{\g\d} \psi_a{}^\g_I (\l^{\d J} \bm g_{-J}{}^I + \bm \l^{\d J} g_{+ J}{}^I ) \non\\
&\quad + \frac{\ri}{2} \eps^{abc} (\g_a)_{\g\d} \psi_b{}^\g_K \psi_c{}^\d_L \, g_+{}^{KP} \bm g_-{}^L{}_P \Big) \ ,
\label{N=4LMA-left}
\end{align}
where the bolded component fields correspond to those of the composite vector multiplet,
\bsubeq
\begin{align}
\bm g_-^{IJ} &= \bm G_-^{IJ}| \ , \quad \bm \l{}_\a^I = \frac{2}{3} \nabla_{\a J} \bm G_-^{IJ}| \ , 
\quad \hat{\bm h}_{+}{}^{IJ} = \frac{\ri}{3} \nabla^{\g [I} \nabla_{\g K} \bm G_-^{J] K}| + 2 w \bm g_{-}{}^{IJ} \ , \\
v_{a} &= e_a{}^m V_{m}| = V_{a}| + \hf \psi_a{}^\a_I V_\a^I | \ , \\
\bm f_{(+) ab} &= 
- \frac{\ri}{24} \eps_{abc} (\g^c)^{\a\b} \nabla_\a^K \nabla_\b^L \bm G_{- KL}| - \hf (\psi_{[a}{}^K \g_{b]} \bm \l_{K}) + \frac{\ri}{2} \psi_a{}^{\g K} \psi_b{}_\g^L \, \bm g_{- KL} \ .
\end{align}
\esubeq

The component anti-self dual vector multiplet action is
\begin{align}
S_{\rm VM}^{(-)} =& - \int \rd^3 x \,e \Big( \eps^{abc} v_{(-) a} \bm f_{(-)bc} + \frac{1}{4} \hat{h}_+{}^{IJ} \bm g_{+ IJ} 
+ \frac{1}{4} \hat{\bm h}_-{}^{IJ} g_{-IJ} - \frac{\ri}{2} \bm \l^{\a I} \l_{\a I} \non\\
&\quad- \frac{1}{2} (\g^a)_{\g\d} \psi_a{}^\g_I (\bm \l^{\d J} g_{-J}{}^I + \l^{\d J} \bm g_{+ J}{}^I ) \non\\
&\quad + \frac{\ri}{2} \eps^{abc} (\g_a)_{\g\d} \psi_b{}^\g_K \psi_c{}^\d_L \, \bm g_+{}^{KP} g_-{}^L{}_P \Big) \ ,
\label{N=4LMA-right}
\end{align}
where
\bsubeq
\begin{align}
\bm g_+^{IJ} &= \bm G_+^{IJ}| \ , \quad \bm \l{}_\a^I = \frac{2}{3} \nabla_{\a J} \bm G_+^{IJ}| \ , \quad \hat{\bm h}_{-}{}^{IJ} = \frac{\ri}{3} \nabla^{\g [I} \nabla_{\g K} \bm G_+^{J] K}| - 2 w \bm g_{+}{}^{IJ} \ , \\
v_{a} &= e_a{}^m V_{m}| = V_{a}| + \hf \psi_a{}^\a_I V_\a^I | \ , \\
\bm f_{(-) ab} 
&= - \frac{\ri}{24} \eps_{abc} (\g^c)^{\a\b} \nabla_\a^K \nabla_\b^L \bm G_{+ KL}| - \hf (\psi_{[a}{}^K \g_{b]} \bm \l_{K}) + \frac{\ri}{2} \psi_a{}^{\g K} \psi_b{}_\g^L \, \bm g_{+ KL} \ .
\end{align}
\esubeq

Plugging in the superspace expressions for $\bm G_{\pm}{}^{IJ}$ one one can construct the 
component fields of the composite 
vector multiplets. The component fields are found to be
{\allowdisplaybreaks
\bsubeq
\bea
\bm g_{\pm}{}^{IJ} &=&
\frac{1}{g_{\pm}} {\hat{h}_{\pm}}^{IJ} 
- \frac{\ri}{2 g_{\pm}^3} \l_{\pm}{}^\a_K \L_{\pm}{}_\a^{[I} g_{\pm}{}^{J] K}
\pm \frac{\ri}{4 g_{\pm}^3} \eps^{IJLP} \l_{\pm}{}^\a_K \l_{\pm}{}_{\a L} g_{\pm}{}_P{}^K
 \ , \\
 \bm \L_{(\pm)}{}_\a^I
 &=&
 \frac{2}{g_{\pm}} \nabla_\a{}^\g \l_{(\pm)}{}_\g^I
 + \frac{2}{g_{\pm}^3} f_{\pm \a\b} \l_{(\pm)}{}^\b_J g_{\pm}{}^{IJ}
 + \frac{1}{3 g_{\pm}^3} \hat{h}_{\mp JK} \l_{(\pm)}{}_\a^I g_{\pm}{}^{JK} \non\\
 && + \frac{2}{3 g_{\pm}^3} \hat{h}_{\mp JK} \l_{(\pm)}{}_\a^J g_{\pm}{}^{KI}
 + \frac{1}{3 g_{\pm}^2} \hat{h}_{\mp}{}^{IJ} \l_{(\pm)}{}_\a^K g_{\pm JK} \non\\
 &&+ \frac{2}{3 g_{\pm}^3} \nabla_{\a\b} g_{\pm JK} \l_{(\pm)}{}^{\b I} g_{\pm}{}^{JK}
 + \frac{4}{3 g_{\pm}^3} \nabla_{\a\b} g_{\pm JK} \l_{(\pm)}{}^{\b J} g_{\pm}{}^{KI} \non\\
 &&- \frac{2}{g_{(\pm)}^3} \nabla_{\a\b} g_{\pm}{}^{IJ} \l_{(\pm)}{}^{\b K} g_{\pm JK} \non\\
 && \pm \frac{1}{g_{\pm}} w \l_{(\pm)}{}_\a^I \pm \frac{8\ri}{g_{\pm}} w_{\a J} g_{\pm}{}^{IJ} 
 + \cO(\l^2)
 \ , \\
\hat{\bm h}_{\pm}{}^{IJ} &=& 
\frac{4}{g_{\pm}} \Box g_{\pm}^{IJ}
+ \frac{2}{g_{\pm}^3} f_{\pm ab} f_{\pm}{}^{ab} g_{\pm}{}^{IJ} 
+ \frac{4}{g_{\pm}^3} \eps^{abc} f_{\pm ab} \nabla_c g_{\pm}{}_K{}^{[I} g_{\pm}{}^{J] K} \non\\
&&- \frac{1}{4 g_{\pm}^3} \hat{h}_{\mp}{}^{KL} \hat{h}_{\mp KL} g_{\pm}{}^{IJ}
- \frac{2}{g_{\pm}^3} g_{\pm}{}^{KL} \nabla_a g_{\pm KL} \nabla^a g_{\pm}{}^{IJ} \non\\
&&
+ \frac{1}{g_{\pm}^3} g_{\pm}{}^{IJ} \nabla_a g_{\pm KL} \nabla^a g_{\pm}{}^{KL} 
- \frac{2}{g_{\pm}} w^2 g_{\pm}{}^{IJ}
\pm \frac{2}{g_{\pm}} y g_{\pm}{}^{IJ} \non\\
&&+\  {\rm fermion \ terms} \ , \\
\bm f_{(\pm) mn} &:=& e_m{}^a e_n{}^b \bm f_{(\pm) ab} \non\\
&=& 
\partial_{[m} \Big(\frac{4}{g_{\pm}} f_{(\pm) n]} - \frac{2}{g_{\pm}} V_{n]}{}^{IJ} g_{\pm IJ} \Big)
- \frac{1}{g_{\pm}^3} \partial_{[m} g_{\pm}{}^{IK} \partial_{n]} g_{\pm}{}^{J}{}_K g_{\pm IJ} \non\\
&&+ \ {\rm fermion \ terms} \ ,
\eea
\esubeq
}
where
\be
f_{(\pm)}{}^m = \hf \eps^{mnp} f_{(\pm)}{}_{np} \ .
\ee
Here we have introduced the following:
\bsubeq
\bea
\nabla_a g_{\pm}{}^{IJ} &:=& 
\cD_a g_{\pm}{}^{IJ}
+ \hf \psi_a{}^{\a [I} \l_{\pm}{}_\a^{J]}
\pm \frac{1}{4} \eps^{IJKL} \psi_a{}^\a_K \l_{\pm \a L} \ , \\
\Box g_{\pm}{}^{IJ} &:=& 
\cD^a \cD_a g_{\pm}{}^{IJ} + \frac{1}{4} \cR g_{\pm}{}^{IJ} 
+ {\rm fermion \ terms} \ ,
\eea
\esubeq
and\footnote{We have denoted the component vector derivative 
$\cD_a$ in the same way as the SU(2) superspace covariant derivative. It should be clear from context 
to which we are referring to.}
\be \cD_a := e_a{}^m \partial_m - \hf \omega_a{}^{bc} M_{bc}
- \hf V_a{}^{IJ} N_{IJ} - b_a \mathbb D \ .
\ee


\subsection{$\cN = 4$ topologically massive supergravity in components}
\label{section4.3}

To simplify our results it is useful to make use of the gauge freedom to impose some gauge condition. 
One can always choose a gauge condition where
\be B_A = 0 \ , \quad G_{\pm} = 1 \ . \label{GaugeCondsVM}
\ee
At the component level these require
\be g_{\pm} = 1 \ , \quad \l_\a^I = 0 \ , \quad b_m = 0 
\ .
\ee
The first gauge condition fixes the dilatation transformations, 
the second fixes the $S$-supersymmetry 
transformations and the third fixes the conformal boosts. For a 
right $G^{ij}$ and left $G^{\bar{i}\bar{j}}$ vector multiplet we can 
use the respective SU(2) symmetry to fix their lowest components 
to a constant. This then gives
\be \nabla_a g_{\pm}{}^{IJ} = 2 V_a{}^{K[I} g_{\pm}{}_K{}^{J]} \ .
\ee

With the above gauge conditons we find
\bsubeq
\bea
\hat{\bm h}_{\pm}^{IJ} g_{\pm IJ} &=& 
2 \cR+ 4 f_{\pm ab} f_{\pm}{}^{ab}
- \hf \hat{h}_{\mp}{}^{IJ} \hat{h}_{\mp IJ} \non\\
&& - 2 V_a{}_{KL} V^a{}^{KL} 
+ 4 V_a{}^{IK} V^a{}^{JL} g_{\pm IJ} g_{\pm KL}
\non\\
&&- 4 w^2 \pm 4 y + {\rm fermion \ terms}
\ , \\
\hat{h}_{\pm}^{IJ} \bm g_{\pm IJ} &=& 
\hat{h}_{\pm}^{IJ} \hat{h}_{\pm IJ}
\ , \\
{\bm f}_{(\pm) mn} &=& 
\partial_{[m} \Big( 
4 f_{(\pm) n]}
- 2 V_{n]}{}^{IJ} g_{\pm IJ}
\Big) 
+ {\rm fermion \ terms} \ .
\eea
\esubeq

Using the above conditions one finds (upon integrating by parts) the self-dual vector multiplet action is
\begin{align}
S_{\rm VM}^{(+)} =& - \int \rd^3 x \,e \Big( \hf \cR - f_{(+) ab} f_{(+)}^{ab} - 2 f_{(+)}^a V_a{}^{IJ} g_{+ IJ}  
- \frac{1}{2} V_a{}_{KL} V^a{}^{KL} \non\\
&+ V_a{}^{IK} V^a{}^{JL} g_{+ IJ} g_{+ KL} + \frac{1}{8} \hat{h}_-^{IJ} \hat{h}_{- IJ}
- w^2 + y
+ {\rm fermion \ terms}
 \Big) \ ,
\end{align}
while the anti-self-dual vector multiplet action is
\begin{align}
S_{\rm VM}^{(-)} =& - \int \rd^3 x \,e \Big( \hf \cR - f_{(-) ab} f_{(-)}^{ab} - 2 f_{(-)}^a V_a{}^{IJ} g_{- IJ}  
- \frac{1}{2} V_a{}_{KL} V^a{}^{KL} \non\\
&+ V_a{}^{IK} V^a{}^{JL} g_{- IJ} g_{- KL} + \frac{1}{8} \hat{h}_+^{IJ} \hat{h}_{+ IJ}
- w^2 - y
+ {\rm fermion \ terms}
 \Big) \ .
\end{align}

The complete component action for minimal $\cN = 4$ topologically massive supergravity (\ref{SUGRA-matteractionN=4}) 
is then given by 
\bea  
\kappa S_{\pm} = \frac{1}{\mu} S_{\rm CSG} + S_{{\rm VM}}^{(\pm)} \ ,
\eea
where $S_{\rm CSG}$ is the component action \eqref{N=4CSGActioN}. As a simple check one 
can readily verify that the 
equation of motion on the field $y$ gives
\be w = \mp \mu \ ,
\ee
which is consistent with the supergravity equation of motion being $W = \mp \mu G_{\pm}$ 
in the presence of the vector multiplet compensator.

For completeness we will also give the component action in isospinor 
notation.  
The $\cN=4$ conformal supergravity action \eqref{N=4CSGActioN} becomes 
\begin{align}
S_{\rm CSG} =& \  \frac{1}{8} \int \rd^3 x \,e \,\Big\{ \eps^{abc} \big( \omega_{a}{}^{fg} \cR_{bc}{}_{fg} - \frac{2}{3} \omega_{af}{}^g \omega_{bg}{}^h \omega_{ch}{}^f 
\non\\
&- 4 {\cR}_{ab}{}^{ij} V_c{}_{ij} - \frac{8}{3} V_a{}_i{}^j V_b{}_j{}^k V_c{}_k{}^i \non\\
&- 4 {\cR}_{ab}{}^{\bar{i}\bar{j}} V_c{}_{\bar{i}\bar{j}} - \frac{8}{3} V_a{}_{\bar{i}}{}^{\bar{j}} V_b{}_{\bar{j}}{}^{\bar{k}} V_c{}_{\bar{k}}{}^{\bar{i}}
 \big)  \non\\
 &- 32 \ri w^\a_{i\bar{i}} w_\a^{i\bar{i}} - 8 w y 
 - 16 \ri \,\psi_a{}^\a_{i\bar{i}} (\g^a)_{\a}{}^\b w_\b^{i\bar{i}} w - 2 \ri \eps^{abc} (\g_a)_{\a\b} \psi_b{}^\a_{i\bar{i}} \psi_c{}^{\b i\bar{i}} w^2 \Big\} \ ,
\end{align}
where the component SU(2) curvatures $\cR_{ab}{}^{ij}$ and $\cR_{ab}{}^{\bar{i}\bar{j}}$ are
\bsubeq
\begin{align}
{\cR}_{ab}{}^{ij} &:= 2 e_a{}^m e_b{}^n \partial_{[m} V_{n]}{}^{ij} - 2 V_{[a}{}{}^{ik} V_{b] k}{}^j
~,
\\
{\cR}_{ab}{}^{\bar{i}\bar{j}} &:= 2 e_a{}^m e_b{}^n \partial_{[m} V_{n]}{}^{\bar{i}\bar{j}} - 2 V_{[a}{}{}^{\bar{i}\bar{k}} V_{b] \bar{k}}{}^{\bar{j}} \ .
\end{align}
\esubeq
The self-dual vector multiplet action in isospinor notation is
\begin{align}
S_{\rm VM}^{(+)} =& - \int \rd^3 x \,e \Big(\hf \cR 
- f_{(+) ab} f_{(+)}^{ab} 
- 4 f_{(+)}^a V_a{}^{\bar{i}\bar{j}} g_{+\bar{i}\bar{j}}  
- V_a{}_{\bar{i}\bar{j}} V^a{}^{\bar{i}\bar{j}} \non\\
&+ 2 V_a{}^{\bar{i}\bar{k}} V^a{}^{\bar{j}\bar{l}} g_{+ \bar{i}\bar{j}} g_{+ \bar{k}\bar{l}} + \frac{1}{4} \hat{h}_-^{ij} \hat{h}_{- ij}
- w^2 + y
+ {\rm fermion \ terms}
 \Big) \ ,
\end{align}
while the anti-self-dual vector multiplet action is
\begin{align}
S_{\rm VM}^{(-)} =& - \int \rd^3 x \,e \Big(\hf \cR - f_{(-) ab} f_{(-)}^{ab} 
- 4 f_{(-)}^a V_a{}^{ij} g_{- ij}  
- V_a{}_{ij} V^a{}^{ij} \non\\
&+ 2 V_a{}^{ik} V^a{}^{jl} g_{- ij} g_{- kl} + \frac{1}{4} \hat{h}_+^{\bar{i}\bar{j}} \hat{h}_{+ \bar{i}\bar{j}}
- w^2 - y
+ {\rm fermion \ terms}\Big) \ .
 \label{428}
\end{align}

Having derived the component actions for minimal $\cN = 4$ topologically massive supergravity, it is worth 
elaborating on these results further. For instance, if we consider just one of 
the vector multiplet 
actions without the conformal supergravity action, one can see that the
equation of motion for $y$ leads to an inconsistency. 
This is equivalent to the fact that the superfield equations of motion for the $\cN=4$
gravitational superfield\footnote{The $\cN=4$
gravitational superfield is a scalar prepotential describing the multiplet
of $\cN=4$ conformal supergravity. It is the 3D $\cN=4$ 
counterpart of the  $\cN=2$ 
gravitational superfield in four dimensions \cite{KT}.}
derived from the actions $S_{\rm VM}^{(+)} $
and $S_{\rm VM}^{(-)} $ are $G_+ =0$ and $G_-=0$, respectively, and these 
equations are inconsistent with the requirements $G_\pm \neq 0$.
However, 
one gets consistent equations of motion
if one adds the left and right vector multiplets  \cite{KN14}
and considers 
the action
\be 
S = S_{\rm VM}^{(+)} + S_{\rm VM}^{(-)}  \ .
\label{429}
\ee
Now the superfield equation of motion for the $\cN=4$
gravitational superfield is \cite{KN14}
\bea
G_+- G_-=0~,
\eea
which is completely consistent. Moreover, this equation is consistent with our gauge conditions 
because imposing the gauge $G_+ = 1$ implies $G_-=1$, 
which in turn implies that the auxiliary field $y$ cancels. 
Furthermore, the fields $w$ and $\hat{h}^{IJ}$ become auxiliary 
and their equation of motion is the requirement that they vanish. The equations of motion on the SU(2) connections 
requires $f_{(-)}{}_a = f_{(+)}{}_a = 0$ and we are left with just the $\cN=4$ Poincar\'e supergravity action (up to a normalisation factor)
\begin{align}
S =& - \int \rd^3 x \,e \,
\cR ~+ ~{\rm fermion \ terms}
  \ .
\end{align}
In the presence of the conformal supergravity action the gauge conditions $G_+ = G_- = 1$ are no longer consistent \cite{KN14} and instead 
one has to use the results in subsection \ref{CompVMactions} in the general gauge. If one also adds to \eqref{429} the supersymmetric cosmological 
term \cite{KLT-M11}, the resulting theory corresponds to (2,2) AdS supergravity
as was described in detail in  \cite{KN14,KLT-M11}.

It is worth mentioning some simplifications that can be made to the $\cN = 4$ topologically massive supergravity actions 
upon using the equations of motion.
To illustrate this let us consider the theory with a self-dual vector multiplet. In this case the equation of motion for the $\rm SU(2)_L$ gauge field is
\be {\cR}_{ab}{}^{ij} = 0 \ ,
\ee
which tells us that 
the $\rm SU(2)_L$ gauge field can be completely gauged away. 
The equation of motion for the 
auxiliary field $\hat{h}^{ij}$ sets the auxiliary field to zero and removes it from the action. The equation of motion on $y$ just sets $w = - \mu$ 
and gives rise to a cosmological term. The resulting action is
\begin{align}
\k S_{+} =& \int \rd^3 x \,e \, \Big[ \frac{1}{8 \mu} \Big\{ \eps^{abc} \big( \omega_{a}{}^{fg} \cR_{bc}{}_{fg} 
- \frac{2}{3} \omega_{af}{}^g \omega_{bg}{}^h \omega_{ch}{}^f 
\non\\
&- 4 {\cR}_{ab}{}^{\bar{i}\bar{j}} V_c{}_{\bar{i}\bar{j}} - \frac{8}{3} V_a{}_{\bar{i}}{}^{\bar{j}} V_b{}_{\bar{j}}{}^{\bar{k}} V_c{}_{\bar{k}}{}^{\bar{i}}
 \big) \Big\} \non\\
  &- \hf \cR + \mu^2
+ f_{(+) ab} f_{(+)}^{ab} 
+ 4 f_{(+)}^a V_a{}^{\bar{i}\bar{j}} g_{+\bar{i}\bar{j}}  
+ V_a{}_{\bar{i}\bar{j}} V^a{}^{\bar{i}\bar{j}} \non\\
&- 2 V_a{}^{\bar{i}\bar{k}} V^a{}^{\bar{j}\bar{l}} g_{+ \bar{i}\bar{j}} g_{+ \bar{k}\bar{l}}
+ {\rm fermion \ terms} \Big] \ .
\end{align}
Similar simplifications can be made for the anti-self dual vector multiplet action.

We can now show how to derive the supergravity action postulated in \cite{LS16}
from our theory $S_-$. The crucial observation is that the U(1) gauge field
appears in the action \eqref{428} only via its field strength $ f_{(-) ab}$, and therefore
it may be dualised into a scalar field.  To implement this, we replace 
\eqref{428} with an equivalent first-order action
\begin{align}
S^{(-)}_{\rm FO} =& - \int \rd^3 x \,e \Big( \hf \cR - f_{(-) ab} f_{(-)}^{ab} - 4 f_{(-)}^a V_a{}^{ij} g_{ij}  
- V_a{}_{ij} V^a{}^{ij} + 2  V_a{}^{ik} V^a{}^{jl} g_{+ ij} g_{+ kl} 
\non\\ 
&+ \frac{1}{4} \hat{h}_+^{\bar{i}\bar{j}} \hat{h}_{+ \bar{i}\bar{j}}
- w^2 - y + 2   f_{(-)}^a \cD_a \varphi
+ {\rm fermion \ terms}
 \Big) \ ,
 \label{434}
\end{align}
where $f_{(-) ab }$ is  an unconstrained antisymmetric tensor field, 
and $\vf$ a Lagrange multipler. 
Varying $\vf$ gives $\cD_a f^a_{(-)} =0$, and therefore $f_{(-) ab }$ 
becomes the field strength of a U(1) vector multiplet.
Then $S^{(-)}_{\rm FO} $ turns into the original action \eqref{428}.
On the other hand, we may integrate out $f_{(-) ab }$ from $S^{(-)}_{\rm FO} $
using its equation of motion 
\be f_{(-) a} = V_a{}^{ij} g_{ij} - \hf \cD_a \varphi \ .
\ee
Plugging this back into \eqref{434} gives the dual action
\bea
S^{(-)}_{\rm hyper}
 &=& - \int \rd^3 x \,e \Big( \hf \cR  
 - \hf \cD^a \varphi \cD_a \varphi
+ 2 \cD_a \varphi V^{a ij} g_{ij}
- 2 V_a{}_{ij} V^a{}^{ij} \non\\
&& + \frac{1}{4} \hat{h}_+^{\bar{i}\bar{j}} \hat{h}_{+ \bar{i}\bar{j}}
- w^2 - y
+ {\rm fermion \ terms}
 \Big) \ ,
\eea
where we used
\be V_a{}^{ik} V^{a jl} g_{ij} g_{kl}
= V_a^{ij} V^{a kl} g_{ij} g_{kl}
- \hf V_{a ij} V^{a ij} \ .
\ee
If we impose a Weyl gauge $\varphi = 1$ and make use of the equation of motion 
for the  auxiliary field $\hat{h}_+^{\bar{i}\bar{j}} $, which is 
$\hat{h}_+^{\bar{i}\bar{j}} = 0$, 
we recover the bosonic matter sector of the topologically massive supergravity action in \cite{LS16} 
up to conventions and fermion terms. 
Since the auxiliary field  $\hat{h}_+^{\bar{i}\bar{j}} $ has been integrated out, 
the action given in \cite{LS16} does not appear to be off-shell.



\section{Discussion} \label{discussion}

In this paper we constructed minimal $\cN=4$ topologically massive supergravity.
It has several unique features that we summarise here.
\begin{itemize}

\item Unlike the other 
$\cN$-extended TMSG theories with $\cN \leq 4$
\cite{DK,Deser,KLRST-M,KN14},
its action cannot be viewed as the supergravity action
(with or without a supersymmetric cosmological term) augmented by 
the conformal supergravity action playing the role of a topological mass term.
The point is that the theory becomes inconsistent upon removing the conformal supergravity action, 
as was explained in section \ref{section4.3}.

\item Our theory makes use of a single superconformal compensator.
We recall that all known Poincar\'e or AdS supergravity theories with eight supercharges
in diverse dimensions require, in general,  two such compensators 
in order for the corresponding dynamics to be consistent. 
One known exception is the off-shell formulation for 4D $\cN=2$ 
AdS supergravity given in \cite{BK11}, which makes use a single massive tensor compensator (described by an unconstrained chiral scalar prepotential)
 and no compensating vector multiplet.\footnote{The vector multiplet has been eaten up by the tensor multiplet. 
The vector compensator acts as a St\"uckelberg field to give mass 
to the tensor multiplet. This is an example of the phenomenon observed originally in \cite{LM} and studied in detail 
in \cite{K08,Dall'Agata:2003yr,DSV,DF,LS,Theis:2004pa,K-tensor}.}
In the case of higher derivative theories, two compensators are no longer required.
This was observed in four dimensions for models  
involving the $\cN=2$ supersymmetric $R^2$ term 
\cite{KN15}, and in three dimensions for $\cN=4$ topologically massive 
supergravity \cite{LS16}.

\item Our minimal TMSG theory does not allow any supersymmetric cosmological term.
However, a cosmological term gets generated at the component level upon 
integrating out the auxiliary fields. This is manifested in the fact that the critical 
(4,0) AdS superspace \cite{KLT-M12} is a maximally supersymmetric solution of the theory.

\item The theory has only one coupling constant.

\item Our minimal TMSG theory is the first off-shell $\cN=4$ supergravity theory in three dimensions with the property that 
the critical (4,0) AdS superspace \cite{KLT-M12}
is a  solution of the theory. Upon integrating out the auxiliary fields we recover the model discussed in \cite{LS16}.

\item Our theory is an off-shell $\cN=4$ supersymmetric extension of 
chiral gravity  \cite{LSS}. It is obvious that such an extension, which has never been constructed before,  must involve a single conformal compensator.
\end{itemize}
The above features demonstrate the physical relevance of the theory 
proposed.

As mentioned in section 1, there exist $\cN=6$ and $\cN=8$ supersymmetric 
extensions of chiral gravity \cite{LSS}. Unlike our  theory, 
these TMSG theories are necessarily on-shell. The off-shell structure 
of our $\cN=4$ theory is indispensable for at least two reasons:
(i) it allows for the general coupling to matter supermultiplets; and (ii) 
at the quantum level, it allows one to derive supersymmetric power-counting rules
through the use of supergraph techniques. 

In the on-shell construction of topologically gauged $\cN=6 $ and $\cN=8$  ABJM type theories
 \cite{Chu:2009gi,Gran:2012mg,Nilsson:2013fya},    
a crucial role is played by a  sixth order scalar potential. 
In the off-shell approach,  such a scalar potential 
is automatically generated upon elimination of the auxiliary fields, 
as was demonstrated in \cite{BILPSZ} where  the  $\cN=6 $ and $\cN=8$ ABJM models 
were realised in $\cN=3$ harmonic superspace. There is an analogous feature in 
our actions. Specifically, before imposing any gauge condition there is a term $w^2 g_{\pm}$ in our actions  
and upon eliminating the auxiliary fields the term $\mu^2 g_{\pm}^3$ is generated. This term plays 
a similar role as the sixth order polynomial in \cite{Chu:2009gi,Gran:2012mg,Nilsson:2013fya} in the 
sense that its coefficient is fixed by the equations of motion (in terms of the coupling coefficient of the conformal supergravity action) 
and the conformal coupling between the Einstein-Hilbert term and the $\cO(2)$ multiplet. 
In this respect our model is akin to those of \cite{Chu:2009gi,Gran:2012mg,Nilsson:2013fya}.

Both models for minimal $\cN=4$ topologically massive supergravity 
constructed in this paper possess  dual formulations. 
They are obtained  by replacing the vector multiplet actions $S^{(+)}_{\rm VM}$
and $S^{(-)}_{\rm VM}$ with off-shell hypermultiplet actions
$S^{(+)}_{\rm HM}$ and $S^{(-)}_{\rm HM}$, respectively, such that
\bea
 S^{(+)}_{{\rm HM}} :=  -\frac{\ri}{2\pi} \oint (v_{\rm R} , \rd v_{\rm R}) \int 
  \rd^{3|8} z
 \, E\,C_{\rm R}^{(-4)}  \Upsilon^{(1)}_\rR \breve\Upsilon^{(1)}_\rR  ~, 
\eea
and similarly for the left hypermultiplet action $S^{(-)}_{\rm HM}$. 
In the dual formulation, its compensating multiplet 
is the so-called  polar hypermultiplet described by the weight-one arctic multiplet 
$\Upsilon^{(1)}_\rR $ and its smile conjugate
 $\breve\Upsilon^{(1)}_\rR $. Duality between the theories with actions 
 $S^{(+)}_{\rm VM}$
 and $S^{(+)}_{\rm HM}$ can be shown in complete analogy with the 4D $\cN=2$ 
 case \cite{K08}.

$~$\\
\noindent
{\bf Acknowledgements:}\\
SMK acknowledges the hospitality of the
Arnold Sommerfeld Center for Theoretical Physics at the Ludwig Maximilian University 
of Munich in July 2015, and of the Theoretical Physics Group at Imperial College, 
London in April 2016.
SMK and JN thank the Galileo Galilei Institute for Theoretical Physics for the hospitality 
and the INFN for partial support during the completion of this work in September 2016.
The work of SMK was supported in part by the Australian 
Research Council,  project No. DP140103925. 
JN acknowledges support from GIF -- the German-Israeli Foundation 
for Scientific Research and Development. I.S. would like to thank DAMTP at Cambridge University for hospitality during the initial stages of this work. I.S. was supported by the DFG Transregional Collaborative Research Centre TRR 33 and the DFG cluster of excellence ``Origin and Structure of the Universe''.


\appendix



\section{The geometry of $\cN = 4$ conformal superspace} \label{geometry}

Here we collect the essential details of the $\cN = 4$ 
superspace geometry of \cite{BKNT-M1}. 
We refer the reader to \cite{KLT-M11,BKNT-M1} for our conventions for 3D spinors.

We begin with a curved three-dimensional $\cN=4$ superspace
 $\cM^{3|8}$ parametrized by
local bosonic $(x^m)$ and fermionic coordinates $(\theta^\m_I)$:
\be z^M = (x^m, \ \q^\mu_I) \ ,
\ee
where $m = 0, 1, 2$, $\mu = 1, 2$ and $I = 1, \cdots , 4$. The 
structure group is chosen to be ${\rm OSp}(4|4, {\mathbb R})$ and 
the covariant derivatives are postulated to have the form
\be
\nabla_A = 
E_A
- \o_A{}^{\underline b} X_{\underline b} 
= 
E_A
- \hf \Omega_A{}^{bc} M_{bc} - \hf \Phi_A{}^{PQ} N_{PQ} - B_A \mathbb D - \mathfrak{F}_A{}^B K_B \ .
\label{A.2}
\ee
Here $E_A = E_A{}^M \partial_M$ is the inverse vielbein, 
$M_{ab}$ are the Lorentz generators, $N_{IJ}$ are generators of the 
$\rm SO(4)$ group, $\mathbb D$ is the dilatation generator and $K_A = (K_a , S_\a^I)$ are the special superconformal 
generators.

The Lorentz generators obey
\bsubeq \label{SCA}
\begin{align}
[M_{ab} , M_{cd}] &= 2 \eta_{c[a} M_{b] d} - 2 \eta_{d [a} M_{b] c} \ , \\
[M_{ab} , \nabla_c ] &= 2 \eta_{c [a} \nabla_{b]} \ , \quad [M_{\a\b} , \nabla_\g^I] = \eps_{\g(\a} \nabla_{\b)}^I \ .
\end{align}
The $\rm SO(4)$ and dilatation generators obey
\begin{align}
[N_{KL} , N^{IJ}] &= 2 \d^I_{[K} N_{L]}{}^J - 2 \d^J_{[K} N_{L]}{}^I \ , \quad [N_{KL} , \nabla_\a^I] = 2 \d^I_{[K} \nabla_{\a L]} \ ,  \\
[\mathbb D , \nabla_a] &= \nabla_a \ , \quad [\mathbb D , \nabla_\a^I] = \hf \nabla_\a^I \ .
\end{align}
The Lorentz and $\rm SO(4)$ 
generators
act
on the special conformal generators $K_A$ as
\begin{align}
[M_{ab} , K_c] &= 2 \eta_{c[a} K_{b]} \ , \quad [M_{\a\b} , S_\g^I] = \eps_{\g(\a} S_{\b)}^I \ , \\
[N_{KL} , S_\a^I] &= 2 \d^I_{[K} S_{\a L]} \ ,
\end{align}
while the dilatation generator acts on  $K_A$ as
\begin{align}
[\mathbb D , K_a] = - K_a \ , \quad [\mathbb D, S_\a^I] &= - \hf S_\a^I \ .
\end{align}
Among themselves, the generators $K_A$ obey the algebra
\begin{align}
\{ S_\a^I , S_\b^J \} = 2 \ri \d^{IJ} (\g^c)_{\a\b} K_c \ .
\end{align}
Finally, the algebra of $K_A$ with $\nabla_A$ is given by
\begin{align}
[K_a , \nabla_b] &= 2 \eta_{ab} \mathbb D + 2 M_{ab} \ , \\
[K_a , \nabla_\a^I ] &= - \ri (\g_a)_\a{}^\b S_\b^I \ , \\
[S_\a^I , \nabla_a] &= \ri (\g_a)_\a{}^\b \nabla_{\b}^I \ , \\
\{ S_\a^I , \nabla_\b^J \} &= 2 \eps_{\a\b} \d^{IJ} \mathbb D - 2 \d^{IJ} M_{\a \b} - 2 \eps_{\a \b} N^{IJ} \ .
\end{align}
\esubeq

The covariant derivatives obey the (anti-)commutation relations of the form
\begin{align}
[ \nabla_A , \nabla_B \}
	&= -T_{AB}{}^C \nabla_C
	- \frac{1}{2} \RM_{AB}{}^{cd} M_{cd}
	- \frac{1}{2} \RN_{AB}{}^{PQ} N_{PQ}
	\non \\ & \quad
	- \RD_{AB} \mathbb D
	- \RS_{AB}{}^\g_K S_\g^K
	- \RK_{AB}{}^c K_c~, \label{nablanabla}
\end{align}
where $T_{AB}{}^C$ is the torsion, and $\RM_{AB}{}^{cd}$, $\RN_{AB}{}^{PQ}$, $\RD_{AB}$, $\RS_{AB}{}^\g_K$ and $\RK_{AB}{}^c$ 
are the curvatures corresponding to the Lorentz, $\rm SO(4)$, dilatation, $S$-supersymmetry and special conformal boosts, 
respectively.

The full gauge group  of conformal supergravity, $\cG$, 
is generated by 
{\it covariant general coordinate transformations}, 
$\delta_{\rm cgct}$, associated with a parameter $\xi^A$ and 
{\it standard superconformal transformations}, 
$\delta_{\cH}$, associated with a parameter $\L^{\ul a}$. 
The latter include
the dilatation,
Lorentz, 
$\rm SO(4)$, 
and special conformal
(bosonic and fermionic) transformations.
The covariant derivatives transform as
\bea 
\d_\cG \nabla_A &=& [\cK , \nabla_A] \ ,
\label{TransCD}
\eea
where $\cK$ denotes the first-order differential operator
\bea
\cK = \xi^C \nabla_C + \hf \L^{ab} M_{ab} + \hf \L^{IJ} N_{IJ} + \L \mathbb D + \L^A K_A ~.
\eea
Covariant (or tensor) superfields transform as
\bea 
\d_{\cG} T &=& \cK T~.
\eea

In order to describe the Weyl multiplet of conformal supergravity, 
some of the components of the torsion and curvatures must be constrained. Following \cite{BKNT-M1}, the 
spinor derivative torsion and curvatures are chosen to resemble super-Yang Mills
\be \{ \nabla_\a^I , \nabla_\b^J \} =  -2 \ri \ve_{\a\b} \cW^{IJ} \ ,
\ee
where $\cW^{IJ}$ is some operator that takes values 
in the superconformal algebra, with $P_A$ replaced by $\nabla_A$. In 
\cite{BKNT-M1} it was shown how to constrain $W^{IJ}$ entirely in terms of the super Cotton tensor (or scalar for $\cN = 4$). 
The super Cotton scalar $W$,
is a primary superfield of dimension 1,
\be
S_\alpha^I W = 0~, \quad K_a W =0 \ , \quad
\mathbb D W = W \ .
\ee
The algebra of covariant derivatives is
\begin{subequations} \label{covDN>3}
\begin{align} 
\{ \nabla_\a^I , \nabla_\b^J \} &= 2 \ri \d^{IJ} \nabla_{\a\b} + \ri \eps_{\a\b} \eps^{IJKL} W N_{KL} 
- \ri \eps_{\a\b} \eps^{IJKL} (\nabla^\g_K W) S_{\g L} \non\\
&\qquad +\frac{1}{4} \eps_{\a\b} (\g^c)^{\g\d} \eps^{IJKL} (\nabla_{\g K} \nabla_{\d L} W) K_c \ , \\
[\nabla_a , \nabla_\b^J ] &= \frac{1}{2} \eps^{JPQK} (\g_a)_{\b\g} (\nabla^\g_K W) N_{PQ} \non\\
&\qquad - \frac{1}{4} (\g_a)_{\b\g} \eps^{JKLP} (\nabla^\g_L \nabla^\d_P W) S_{\d K} \non\\
&\qquad - \frac{\ri}{24} (\g_a)_{\b\g} (\g^c)_{\d\rho} \eps^{JKLP} (\nabla^\g_K \nabla^\d_L \nabla^\rho_P W) K_c \ , \\
[\nabla_a , \nabla_b] &
=    \frac{1}{8}   \eps_{abc} (\g^c)_{\a\b} \eps^{PQIJ}
\Big( \ri (\nabla^\a_I \nabla^\b_J W) N_{PQ} \non\\
&\qquad + \frac{\ri}{ 3} \eps^{LIJK} (\nabla^\a_I \nabla^\b_J \nabla^\g_K W) S_{\g L} \non\\
&\qquad + \frac{1}{24} (\g^d)_{\g\d} \eps^{IJKL} (\nabla^\a_I \nabla^\b_J \nabla^\g_K \nabla^\d_L W) K_d \Big) \ ,
\end{align}
\end{subequations}
where the super Cotton scalar $W$ satisfies the following dimension 2 Bianchi identity
\bea
\nabla^{\a I}\nabla_{\a}^JW=\frac{1}{4}\d^{IJ}\nabla^{\a}_P\nabla_{\a}^PW~. \label{2.39}
\eea

For each $\rm SO(4)$ vector $V_I$ we can associate a second-rank isospinor $V_{i\bar{i}}$
\be V_I \leftrightarrow V_{i \bar{i}} := (\t^I)_{i \bar{i}} V_{i \bar{i}} \ , \quad (V_{i \bar{i}})^* = V^{i \bar{i}} \ .
\ee

The original $\rm SO(4)$ connection turns into a sum of two $\rm SU(2)$ connections
\be \Phi_A = (\Phi_{\rm L})_A + (\Phi_{\rm R})_A \ , \quad (\Phi_{\rm L})_A = \Phi_A{}^{kl}  {L}_{kl} \ , \quad (\Phi_{\rm R})_A = \Phi_A{}^{\bar{k} \bar{l}}  { R}_{kl} \ .
\ee
Here $L_{kl}$ is the $ {\rm SU}(2)_{\rL}$ generator 
and  $R_{\bak\bal}$ is the $ {\rm SU}(2)_{\rR}$ generator.
They are related to the SO(4) generators 
$N_{KL}$ as
\bea
N_{KL}~\to~N_{k\bak l\bal}=\ve_{\bak\bal} L_{kl}+\ve_{kl} R_{\bak\bal}~.
\eea
The left and right operators act on the covariant derivatives as
\be [ { L}^{kl} ,  \nabla_\a^{i\bar{i}}] = \eps^{i (k} \nabla_\a^{l) \bar{i}} \ , 
\quad [ {R}^{kl} , \nabla_\a^{i\bar{i}} ] = \eps^{\bar{i}( \bar{k}} \nabla_\a^{i \bar{l})} \ .
\ee
In the isospinor notation, 
the Bianchi identity on $W$ becomes
\be \nabla^{\a i \bar{i}} \nabla_\a^{j \bar{j}} W
= \frac{1}{4} \eps^{ij} \eps^{\bar{i} \bar{j}} 
\nabla^\a_{k \bar{k}} \nabla_\a^{k \bar{k}} W \ .
\ee

The algebra of spinor covariant derivatives becomes
\begin{align}
\{ \nabla_\a^{i \bar{i}} , \nabla_\b^{j \bar{j}}\} &= 2 \ri \eps^{ij} \eps^{\bar{i} \bar{j}} \nabla_{\a\b} 
+ 2 \ri \eps_{\a\b} \eps^{\bar{i} \bar{j}} W  { L}^{ij} 
- 2 \ri \eps_{\a\b} \eps^{ij} W  {R}^{\bar{i} \bar{j}} \non\\
&\quad- \ri \eps_{\a\b} \eps^{ij} \nabla^\g{}_k{}^{\bar{i}} W S_\g^{k \bar{j}} + \ri \eps_{\a\b} \eps^{\bar{i} \bar{j}} \nabla^\g{}^i{}_{\bar{k}} W S_\g^{j \bar{k}} \non\\
&\quad + \frac{1}{4} \eps_{\a\b} \Big( \eps^{ij} \nabla_\g{}_k{}^{\bar{i}} \nabla_\d^{k \bar{j}} W 
- \eps^{\bar{i} \bar{j}} \nabla_\g{}^j{}_{\bar{k}} \nabla_\d^{i \bar{k}} W \Big) K^{\g\d}
\label{A.33}
\end{align}
and the action of the $S$-supersymmetry generator on $\nabla_\a^{i\bar{i}}$ is
\be \{ S_\a^{i \bar{i}} , \nabla_\b^{j \bar{j}} \} = 2 \eps_{\a\b} \eps^{ij} \eps^{\bar{i}\bar{j}} \mathbb{D} - 2 \eps^{ij} \eps^{\bar{i}\bar{j}} M_{\a\b} 
+ 2 \eps_{\a\b} \eps^{\bar{i} \bar{j}} L^{ij} + 2 \eps_{\a\b} \eps^{ij} R^{\bar{i} \bar{j}} \ .
\ee


\section{The geometry of SO(4) superspace} \label{geometrySO4}

For many applications it is useful to work with a superspace formulation with a 
smaller structure group than that of conformal superspace. 
The superspace formulation of \cite{KLT-M11,HIPT}, known as SO(4) superspace, 
provides such a formulation and may be obtained from conformal superspace 
via a degauging procedure \cite{BKNT-M1}. For the $\cN = 4$ case one chooses the 
structure group to be ${\rm SO} (4)$.
The SO(4) superspace formulation for $\cN=4$ conformal supergravity has 
been used to construct general off-shell supergravity-matter couplings \cite{KLT-M11}.

The covariant derivatives have the form:
\bea
\cD_{A} = 
E_{A}
-\O_A
-\F_A
~.
\eea
Here $E_A=E_A{}^M(z)\pa_M$ is the supervielbein, with $\pa_M=\pa/\pa z^M$, 
$\Omega_A$ is the Lorentz connection, and $\F_A=\hf\F_A{}^{KL}N_{KL}$ 
is the SO(4)-connection.
The supergravity gauge group is generated by local transformations of the form
\bea
\d_K\cD_A=[K,\cD_A]~,~~~~~~
K=K^C(z)\cD_C+\hf K^{cd}(z)M_{cd}+\hf K^{PQ}(z)N_{PQ}
~,
\label{SUGRA-gauge-group1}
\eea
with all the gauge parameters obeying natural reality conditions.

The covariant derivatives satisfy the (anti)commutation relations
\bea
{[}\cD_{{A}},\cD_{{B}}\}&=&- T_{{A}{B}}{}^{{C}}\cD_{{C}}
-\hf R_{AB}{}^{KL} N_{KL}
-\hf R_{{A}{B}}{}^{cd} M_{cd}
~,
\label{algebra-4-2-N}
\eea
with  $T_{AB}{}^C$ the torsion,  $R_{AB}{}^{cd}$ the Lorentz curvature and
 $R_{AB}{}^{KL}$ the SO(4) curvature.
 The algebra of covariant derivatives  must be constrained to describe conformal supergravity. The  appropriate constraints \cite{HIPT}
 lead to the following anti-commutation relation \cite{KLT-M11}:
\bsubeq
\bea
\{\cD_\a^I,\cD_\b^J\}&=&
2\ri\d^{IJ}(\g^c)_{\a\b}\cD_c
-2\ri\ve_{\a\b}C^{\g\d}{}^{IJ}M_{\g\d}
-4\ri S^{IJ} M_{\a\b}
\non\\
&&
+\Big(
\ri\ve_{\a\b}W^{IJKL}
-4\ri\ve_{\a\b}S^{K}{}^{[I}\d^{J]L}
+\ri C_{\a\b}{}^{KL}\d^{IJ}
-4\ri C_{\a\b}{}^{K(I}\d^{J)L}
\Big)
N_{KL}
~.~~~~~~~~~
\label{alg-1}
\eea
\esubeq
Here the dimension-1 components are real and satisfy the symmetry properties
\bea
W^{IJKL}=W^{[IJKL]} = \eps^{IJKL} W~,~~~~
S^{IJ}=S^{(IJ)}
~,~~~~
C_a{}^{IJ}=C_a{}^{[IJ]}
~.
\eea
It is useful to decompose the torsion superfield $S^{IJ}$ into its 
trace ($\cS$) and traceless ($\cS^{IJ}$) parts as
\bea
S^{IJ}=\cS\d^{IJ}+\cS^{IJ}~,\quad 
\cS=\frac{1}{\cN}\d_{IJ}S^{IJ}~,\quad
\d_{IJ}\cS^{IJ}=0~.
\eea
The torsion superfields satisfy the Bianchi identities
\bsubeq
\bea
\cD_\a^{I} \cS^{JK}&=&
2\cT_{\a}{}^{I(JK)}
+\cS_\a{}^{(J}\d^{K)I}
-{1\over \cN}\cS_\a{}^{I}\d^{JK}~,
\label{22.18a} \\
\cD_{\a}^{I} C_{\b\g}{}^{JK}
&=&
{2\over 3}\ve_{\a(\b}\Big(
C_{\g)}{}^{IJK}
+3\cT_{\g)}{}^{JKI}
+4(\cD_{\g)}^{[J} \cS)\d^{K]I}
+{(\cN-4)\over \cN}\cS_{\g)}{}^{[J}\d^{K]I}
\Big)
\non\\
&&
+C_{\a\b\g}{}^{IJK}
-2 C_{\a\b\g}{}^{[J}\d^{K]I}
~, \\
0 &=& \Big( \cD^{\g (I} \cD_\g^{J)} - \frac{1}{4} \d^{IJ} \cD^{\g K} \cD_{\g K} - 4 \ri \cS^{IJ}
\Big) W 
~.
\eea
\esubeq

It is often useful to make use of the isomorphism 
 ${\rm SO}(4) \cong  \big( {\rm SU}(2)_{\rL}\times {\rm SU}(2)_{\rR}\big)/{\mathbb Z}_2$
and make use of isospinor notation, $\cD^I_\a \to \cD^{i\bar i}_\a$, by replacing each SO(4) vector 
index  by a pair of isospinor ones. For our notation and conventions we refer the reader to \cite{KLT-M11}.

After introducing isospinor notation, 
the covariant derivatives are
\bea
\cD_{A}&=& (\cD_a, \cD^{i\bar i}_\a )=E_{A}
-\O_{A}
-\F_A
~,
\eea
where the original SO(4) connection $\F_A$ now turns into a sum of  two SU(2) connections
\bea
\F_A=(\F_{\rL})_A+(\F_{\rR})_A~,~~~~
(\F_{\rm L})_A=\Phi_{A}{}^{kl} L_{kl}~,~~
(\F_{\rm R})_A=\Phi_{A}{}^{\bak\bal} R_{\bak\bal}
~.
\eea
The two SU(2) generators act on 
the spinor covariant derivatives $\cD_\a^{i\bai}:=\cD_\a^I(\t_I)^{i\bai}$ as follows:
\bea
&&
{\big [} {L}{}^{kl},\cD_{\a}^{i\bai}{\Big]} =\ve^{i(k} \cD_{\a}^{ l)\bai}
~,~~~
{\big [} {R}{}^{\bak\bal},\cD_{\a}^{i\bai}{\Big]} =\ve^{\bai(\bak} \cD_{\a}^{i \bal)}
~.
\label{acL-R}
\eea

The algebra of spinor covariant derivatives is
\bea
\{\cD_\a^{i\bai},\cD_\b^{j\baj }\}&=& \phantom{+}
2\ri\ve^{ij}\ve^{\bai \baj }(\g^c)_{\a\b}\cD_c
+{2\ri}\ve_{\a\b}\ve^{\bai \baj }(2\cS+X)L^{ij}
-2\ri\ve_{\a\b}\ve^{ij}\cS^{kl}{}^{\bai \baj }L_{kl}
+4\ri C_{\a\b}{}^{\bai \baj }L^{ij}
\non\\
&&
+2\ri\ve_{\a\b}\ve^{ij}(2\cS-X)R^{\bai \baj }
-2\ri\ve_{\a\b}\ve^{\bai \baj }\cS^{ij}{}^{\bak\bal}R_{\bak\bal}
+4\ri C_{\a\b}{}^{ij}R^{\bai \baj }
\non\\
&&
+2\ri\ve_{\a\b}(\ve^{\bai \baj }C^{\g\d}{}^{ij}+\ve^{ij}C^{\g\d}{}^{\bai \baj })M_{\g\d}
-4\ri(\cS^{ij}{}^{\bai \baj }+\ve^{ij}\ve^{\bai \baj }\cS)M_{\a\b}
~,
\label{N=4alg}
\eea
where the torsion components satisfy certain Bianchi identities given 
in \cite{KLT-M11}.\footnote{As compared to \cite{KLT-M11}, we have relabelled the superfield $B_{\a\b}{}^{ij}$ by $C_{\a\b}{}^{ij}$.}


\section{Super-Weyl gauge conditions} \label{2VMtoCSG}

In this appendix we show how one can use the super-Weyl freedom to impose certain gauge conditions in 
SO(4) superspace. In particular, within the $\rm SO(4)$ superspace formulation 
we will show that one can impose either
\be C_a{}^{\bar{i}\bar{j}} = 0 \ , \quad 2 S + W = 0 \label{G+GCond}
\ee
or
\be C_a{}^{ij} = 0 \ , \quad 2 S - W = 0 \ . \label{G-GCond}
\ee

We begin by introducing, within the SO(4) 
superspace geometry, an  
off-shell self-dual vector multiplet $G^{\bar{i}\bar{j}}$ and an anti-self-dual vector multiplet $G^{ij}$. 
They are constrained by the differential constraints for $\cO(2)$ multiplets
\be \cD_\a^{i(\bar{i}} G^{\bar{j}\bar{k})} = 0 \ , \quad \cD_\a^{(i \bar{i}} G^{jk)} = 0 \ .
\ee
Using these constraints it is possible to build some of the components of the torsion in terms of 
these multiplets. In particular, one finds
\bsubeq
\bea
2 S - W &=& \frac{\ri G_+}{8} \cD^{\g i\bar{i}} \cD_{\g i\bar{i}} G_+^{-1} \ , \\
2 S + W &=& \frac{\ri G_-}{8} \cD^{\g i \bar{i}} \cD_{\g i\bar{i}} G_-^{-1} \ , \\
C_{\a\b}{}^{ij} &=& - \frac{\ri}{4} G_+ \cD_\a^{(i \bar{k}} \cD_\b{}^{j)}{}_{\bar{k}} G_+^{-1} \ , \\
C_{\a\b}{}^{\bar{i}\bar{j}} &=& - \frac{\ri}{4} G_- \cD_\a{}^{k(\bar{i}} \cD_\b{}_k{}^{\bar{j})} G_-^{-1} \ , \\
S^{(k}{}_p{}^{\bar{i}\bar{j}} G^{l) p} &=& - \frac{\ri}{16} \{ \cD^{\g p (\bar{i}} , \cD_\g{}_p{}^{\bar{j})} \} G^{kl} \ , \\
S^{ij}{}_{\bar{p}}{}^{(\bar{k}} G^{\bar{l})\bar{p}} &=& - \frac{\ri}{16} \{ \cD^{\g (i \bar{p}} , \cD_\g{}^{j)}{}_{\bar{p}} \} G^{\bar{k}\bar{l}} \ ,
\eea
\esubeq
where $G_+^2 = G^{\bar{i}\bar{j}} G_{\bar{i}\bar{j}}$ and $G_-^2 = G^{ij} G_{ij}$.

The vector multiplets transform homogeneously under super-Weyl transformations
\be 
G^{\bar{i}\bar{j}} \rightarrow \re^{\s} G^{\bar{i}\bar{j}} \ , \quad 
G^{ij} \rightarrow \re^{\s} G^{ij} \ ,
\ee
which tells us that the super-Weyl freedom can be completely fixed by imposing the gauge condition $G_+ = 1$ 
or $G_- = 1$. If we impose $G_+ = 1$ we find the conditions \eqref{G+GCond}, while if we impose $G_- = 1$ we find the 
conditions \eqref{G-GCond}. Therefore, these conditions can always be imposed by an appropriate super-Weyl 
transformation.


\begin{footnotesize}

\end{footnotesize}

\end{document}